\documentclass[pra,floatfix,amsmath,subeqn,amsfonts,superscriptaddress,twocolumn]{revtex4}
\usepackage{amsthm,natbib,graphicx}
\usepackage[dvipdf]{hyperref}
\usepackage[dvips]{changebar}
\usepackage[usenames]{color}
\newcommand{\pd}[2]{\frac{\partial #1}{\partial #2}}
\newcommand{\pdd}[2]{\frac{\partial^2 #1}{\partial #2^2}}
\newcommand{\abs}[1]{\left\lvert#1\right\rvert}

\newcommand{\ud}{\mathrm{d}}
\newcommand{\real}{\mathbb{R}}
\newcommand{\eps}{\varepsilon}
\DeclareMathOperator{\sech}{sech}
\DeclareMathOperator{\sgn}{sgn}
\numberwithin{equation}{section}  

\begin{document}
\title{On Dispersive and Classical Shock Waves in Bose-Einstein
  Condensates and Gas Dynamics}
\date{\today}
\author{M. A. \surname{Hoefer}}
\email{hoefer@colorado.edu}
\author{M. J. \surname{Ablowitz}}
\affiliation{Department of Applied Mathematics, University of
 Colorado, Campus Box 526, Boulder, Colorado 80309-0526}
\author{I. \surname{Coddington}}
\author{E. A. \surname{Cornell}}
\altaffiliation{Also at Quantum Physics Division, National Institute
  of Standards and Technology}
\author{P. \surname{Engels}}
\altaffiliation{Present address:  Department of Physics and Astronomy,
  Washington State University, Pullman, WA  99163}
\author{V. \surname{Schweikhard}}
\affiliation{JILA, National Institute of Standards and Technology and
  University of Colorado, and Department of Physics, University of
  Colorado, Boulder, Colorado 80309-0440, USA}
\begin{abstract}
  A Bose-Einstein condensate (BEC) is a quantum fluid that gives rise
  to interesting shock wave nonlinear dynamics.  Experiments depict a
  BEC that exhibits behavior similar to that of a shock wave in a
  compressible gas, eg.\! traveling fronts with steep gradients.
  However, the governing Gross-Pitaevskii (GP) equation that describes
  the mean field of a BEC admits no dissipation hence classical
  dissipative shock solutions do not explain the phenomena.  Instead,
  wave dynamics with small dispersion is considered and it is shown
  that this provides a mechanism for the generation of a dispersive
  shock wave (DSW).  Computations with the GP equation are compared to
  experiment with excellent agreement.  A comparison between a
  canonical 1D dissipative and dispersive shock problem shows
  significant differences in shock structure and shock front speed.
  Numerical results associated with the three dimensional experiment
  show that three and two dimensional approximations are in excellent
  agreement and one dimensional approximations are in good qualitative
  agreement.  Using one dimensional DSW theory it is argued that the
  experimentally observed blast waves may be viewed as dispersive
  shock waves.
\end{abstract}
\maketitle


\section{Introduction}
It is well known that a shock wave in a compressible fluid is
characterized by a steep jump in gas velocity, density, and
temperature across which there is a dissipation of energy due
fundamentally to collisions of particles.  The aim of this article is
to present experimental and numerical evidence of a different type of
shock wave which is generated in a quantum fluid that is a
Bose-Einstein condensate (BEC). In this case, the shock front is
dominated not by dissipation but rather dispersion.  Viewed locally,
these dispersive shock waves (DSWs) with large amplitude oscillations
and two associated speeds bear little resemblance to their classical,
dissipative counterparts.  However, we demonstrate that a direct
comparison is possible when one considers a mean field theory,
corresponding to the \emph{average} of a DSW.

Since extensive theoretical work has been done in the field of
compressible gas dynamics (cf. \cite{CoFr48}), it is important to
relate this work to the ``dispersive gas dynamics'' which BEC
embodies.  The present work contrasts and compares dissipative and
dispersive shock waves through multidimensional numerical simulation
and analytical studies.  We also provide an explanation of what a BEC
shock wave is in the context of the well understood concept of a
classical shock wave in gas dynamics.

Early experiments studying shock-induced dynamics in BEC were reported
in \cite{Hau01} where a slow-light technique was used to produce a
sharp density depression in a BEC.  Direct experimental imaging of BEC
blast waves has been performed in the rotating context and numerical
solutions of the governing Gross-Pitaevskii (GP) equation were used to
describe the wave dynamics \cite{Engels05}.  Theoretical studies of
the zero dissipation limit of classical gas dynamics as applied to BEC
was discussed in \cite{ZaKu03,Da04} where it was shown that a shock
wave could develop.  Subsequently, in \cite{KaGaKr04} the shock wave
in the small dispersion limit of the one dimensional repulsive GP
equation was analyzed using the Whitham averaging method and the
attractive GP equation was analyzed in \cite{AbdullaevKamchatnov05}.

In the present paper, for the first time a comparison between
dissipative and dispersive shock waves is carried out through a
careful investigation of new experiments and theory in one, two, and
three dimensions.

The outline of this work is as follows.  In section
\ref{sec:non-dimens-gross} we give the relevant dynamical equation,
i.e. the GP equation, and we put it in non-dimensional form and give
the associated conservation laws. In section \ref{sec:experiment} we
present new experimental results depicting "blast" waves in a
non-rotating BEC.  In section \ref{sec:numerical-results} we show that
direct three dimensional numerical simulations, with radial symmetry
using the GP equation give excellent agreement with these experiments.
In section \ref{sec:class-disp-shock} an analysis of dissipative and
dispersive shock waves in two types of one-dimensional systems, the
inviscid Burgers' equation and the Euler equations, is provided.  Two
types of limiting behavior for conservation laws, the
\emph{dissipative regularization} (small dissipation limit) and
\emph{dispersive regularization} (small dispersion limit) are
considered.  We then present numerical evidence showing that the three
dimensional and two dimensional calculations agree extremely well
(less than one percent relative difference). It is also found that the
one dimensional approximation of the 3D blast wave experiments is in
good qualitative agreement.  Using one dimensional theory, we explain
why the experimentally observed blast waves may be viewed as DSWs.

\subsection{Gross-Pitaevskii and the Navier-Stokes Equations}
\label{sec:non-dimens-gross}
An analogy between the classical equations of fluid flow, the
Navier-Stokes (NS) equations, and the density, phase equations for the
wave function (order parameter) associated with a BEC is well known
\cite{PeSm02}.  The crucial difference from NS is a dispersive term
that replaces the dissipative term in classical fluid dynamics.

The GP equation models the mean-field dynamics of the BEC wave
function $\Psi$ and has been shown to be an effective approximation in
many situations.  Experiments in rapidly rotating BECs have provided
evidence of dynamics similar to what is often considered to be ``blast
waves''. Moreover, simulations with the Gross-Pitaevskii equation were
compared with experiment giving support to the validity of the GP
equation in such extreme circumstances \cite{Engels05}.

 The dimensional GP equation is  \cite{PeSm02} 
\begin{equation}
  \label{eq:59}
  i \hbar \Psi_t = -\frac{\hbar^2}{2m} \nabla^2 \Psi + V_0 \Psi + NU_0|\Psi|^2
  \Psi,
\end{equation}
with conservation of particle number
\begin{equation}
  \label{eq:16}
  \int_{\real^3} |\Psi|^2 \,\ud^3 x = 1 .
\end{equation}
The coefficient of nonlinearity, $NU_0 = N4\pi \hbar^2 a_s/m$, is
characterized by the inter-particle scattering length $a_s$ (here
positive representing repulsive particles) and the number of condensed
atoms $N$; the other parameters are the atomic mass of the species
considered ($m$) and Planck's constant divided by $2\pi$ ($\hbar$).
The standard confining harmonic potential (trap) is given by
\begin{equation*}
  V_0(x,y,z) = \frac{m}{2} \left(\omega_{\perp}^2 (x^2 + y^2) +
    \omega_z^2 z^2 \right),
\end{equation*}
where $\omega_{\perp}$ and $\omega_z$ are the radial and axial trap
frequencies respectively.  A convenient normalization for our purposes
is to take \cite{BaJaMa03}
{\setlength\arraycolsep{1pt}
\begin{eqnarray*}
  t' &=& \omega_{\perp} t, \quad \vec{x}' = \frac{\vec{x}}{l}, \quad \Psi' =
  l^{\frac{3}{2}} \Psi, \quad
  l = \left(\frac{4\pi \hbar^2 |a_s| N}{m^2 \omega_{\perp}^2}
  \right)^{\frac{1}{5}}.
\end{eqnarray*}}
\!\!After dropping primes, equation \eqref{eq:59} becomes
\begin{equation}
  \label{eq:1}
  i \eps \Psi_t = -\frac{\eps^2}{2} \nabla^2 \Psi + V_0 \Psi + |\Psi|^2
  \Psi ~,
\end{equation}
where
\begin{equation*}
  \eps = \left( \frac{\hbar}{m \omega_{\perp} (4 \pi a_s N)^2}
  \right)^{\frac{1}{5}} \ll 1,
\end{equation*}
with the normalization of the wavefunction (\ref{eq:16}) preserved
and the trap potential
\begin{equation*}
  V_0(r,z) = \frac{1}{2}(r^2 + \alpha_z z^2), \quad r^2 = x^2 + y^2. 
\end{equation*}
The coefficient $\alpha_z = (\omega_z/\omega_{\perp})^2$ represents
the asymmetry in the harmonic trap.   In the experiments considered,
the parameters are: $N = 3.5\cdot 10^6$ particles, $a_s = 5.5$ nm and
$m = 1.45 \cdot 10^{-25}$ kg for the species $^{87}\text{Rb}$,
$(\omega_{\perp},\omega_z) = 2 \pi(8.3,5.3)$ Hz, and $\alpha_z =
2.45$.  This normalization shows that the dispersion is extremely
small, $\eps = 0.012$. 

Conservation of ``mass'' and ``momentum'' for the GP equation
\eqref{eq:1} are,
\begin{equation}
  \label{eq:27}
  \frac{d}{dt} \int_{\real^3} \abs{\Psi}^2 \, \ud \vec{x} = 0, \quad
  \frac{d}{dt} \int_{\real^3} (\Psi^* \nabla \Psi - \Psi \nabla
  \Psi^*) \, \ud \vec{x} = 0.
\end{equation}
Since it will be useful for later discussions, we give the local
conservation laws in the 1D case (see e.g. \cite{JiLeMc94})
\begin{equation}
  \label{eq:19}
  \begin{split}
    \rho_t + (\rho u)_x &= 0 \\[2mm]
    (\rho u)_t + \left( \rho u^2 + \tfrac{1}{2} \rho^2 \right)_x
    &= \frac{\eps^2}{4} (\rho (\log \rho)_{xx} )_x - \rho V_x,
  \end{split}
\end{equation}
where subscripts denote differentiation. The condensate ``density''
$\rho$ and ``velocity'' $u$ are defined by
\begin{equation*}
  \Psi = \sqrt{\rho} e^{i \phi/\eps}, \quad u =  \phi_x.
\end{equation*}

Equations \eqref{eq:19} give an alternative formulation of the GP
equation in terms of ``fluid-like'' variables. Since the $\eps^2$ term
is obtained from the linear dispersive term in eq. \eqref{eq:1}, we
call this the dispersive term.

 The Navier-Stokes equations for a 1D compressible gas with
density $\rho$ and velocity $u$ can be written \cite{LiepmannRoshko57}
\begin{equation}
  \label{eq:35}
  \begin{split}
    \rho_t + (\rho u)_x &= 0 \\[2mm]
    (\rho u)_t + \left( \rho u^2 + P \right)_x
    &= \eps^2 u_{xx} + \rho F,
  \end{split}
\end{equation}
where $P$ is the pressure and $F$ is an external force per unit mass.
The positive coefficient $\eps^2$ represents dissipative effects due
to viscous shear and heat transfer.  If the pressure law
\begin{equation*}
  P = \tfrac{1}{2} \rho^2
\end{equation*}
is assumed (for example, a perfect, isentropic gas with adiabatic
constant $\gamma = 2$ or, equivalently, the shallow water equations
for height $\rho$ and velocity $u$), then the NS equations
\eqref{eq:35} correspond to the GP conservation equations
\eqref{eq:19} when $\eps^2 = 0$ and $F = -V_x$.  Equations
\eqref{eq:35} for the case $\eps^2 = 0$ and $F = 0$ are called the
Euler equations.  To compare different types of shock waves, we are
interested in the dispersive regularization ($\eps^2 \to 0$ in
\eqref{eq:19}) as compared to the dissipative regularization ($\eps^2
\to 0$ in \eqref{eq:35}) of the Euler equations ($V_x = F =
0$).

A remark regarding the form of the above equations.  Many authors use
the velocity form 
\begin{equation*}
  \begin{split}
    \rho_t + (\rho u)_x &= 0 \\[2mm]
    u_t + \left( \tfrac{1}{2} u^2 + \rho \right)_x &= \frac{\eps^2}{4
      \rho} (\rho (\log \rho)_{xx} )_x - V_x,
  \end{split}
\end{equation*}
of equation \eqref{eq:19} by implicitly assuming differentiability and
that $\rho \ne 0$.  As we will be interested in \emph{weak}, hence not
everywhere smooth, solutions to the dissipative regularization of the
Euler equations, it is necessary to maintain the form of the
conservation laws as derived directly from the original integral
formulation \eqref{eq:27}.  It is well known that weak solutions to
different forms of the same conservation law can be quite different.
One must also be careful when dealing with the vacuum state $\rho = 0$
as in classical gas dynamics.  The momentum equation (\ref{eq:19})
takes both of these issues into account.

\section{Experiment}
\label{sec:experiment}
In order to investigate the fundamental nature of shock waves in a
quantum fluid, we have performed new experiments that involve blast
pulses in BECs.  In contrast to the experiments described in
\cite{Engels05}, the experiments analyzed in this work are all done
with non-rotating condensates.  We have succeeded in directly imaging
dispersive shock waves in these systems, and the particular geometry
of these experiments makes them amenable to the theoretical analysis
presented in this paper.

Condensates consisting of approximately 3.5 million Rb atoms were
prepared in an axisymmetric trap with trapping frequencies of
$(\omega_{\perp},\omega_z) = 2\pi(8.3,5.3)$ Hz; $\omega_{\perp}$ is
the radial frequency and $\omega_z$ is the axial frequency.  After the
condensate was formed, a short, tightly focused laser beam was pulsed
along the $z$-axis through the center of the BEC.  The wavelength of
the laser was 660 nm, which is far red-detuned from the Rb
transitions.  The pulse rapidly pushes atoms from the center of the
BEC radially outward, leading to the formation of a density ring.
Before imaging, an anti-trapping technique was used to enlarge the
features of the blast wave.  In brief, a rapid expansion of the BEC is
created by changing the internal state of the atoms such that they are
radially expelled by the strong magnetic fields forming the trap.
Details about the anti-trapped expansion are described in
\cite{Coddington04}.  While the anti-trapped expansion changes the
scale of the features involved, it does not alter the qualitative
appearance of the shock phenomena, as is confirmed by our numerical
simulations.

\begin{figure}
  \centering
  \begin{tabular}{cc}
    \begin{minipage}{42mm}
      \includegraphics[width=42mm]{
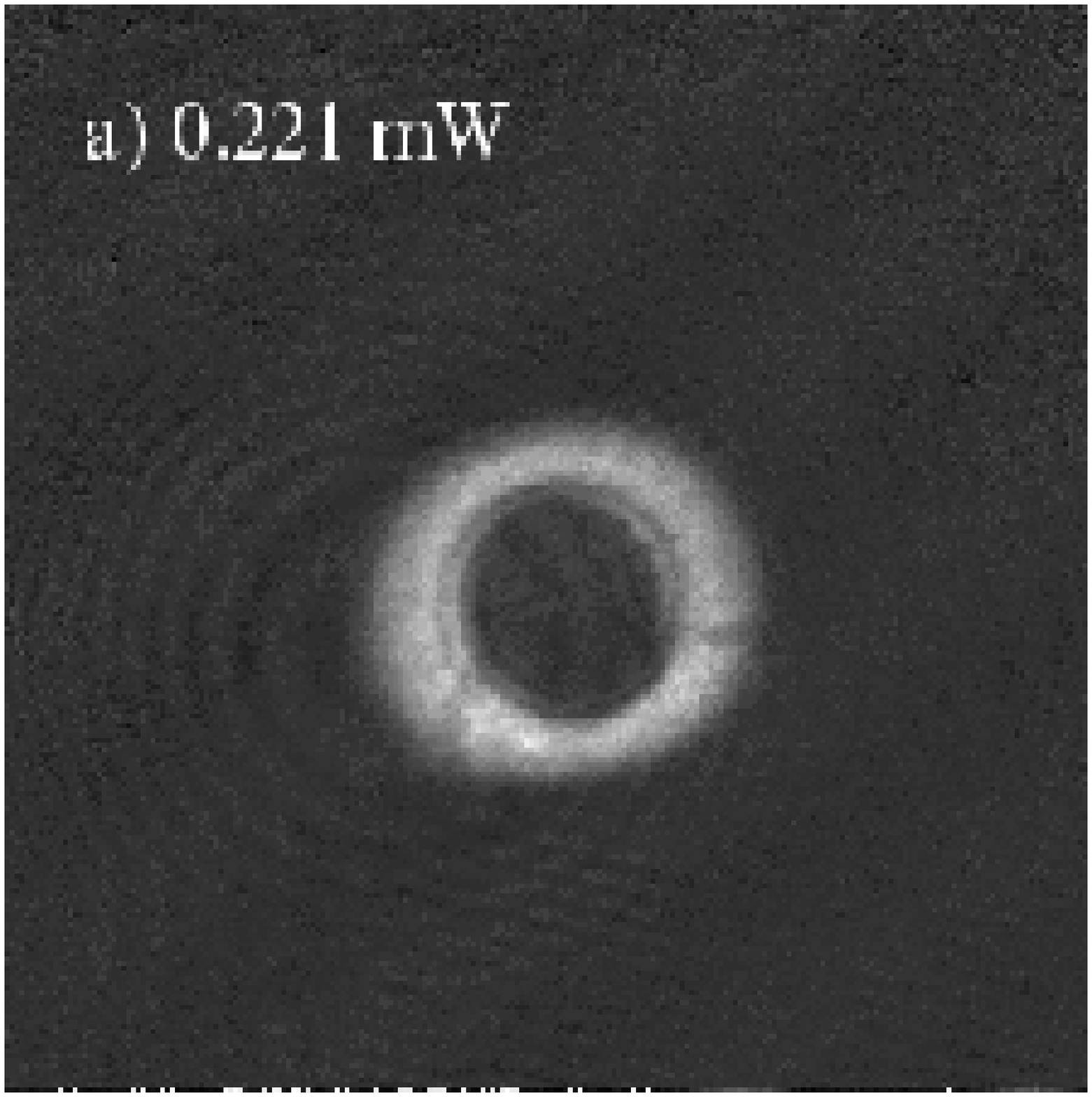}
    \end{minipage}
    &
    \begin{minipage}{42mm}
      \includegraphics[width=42mm]{
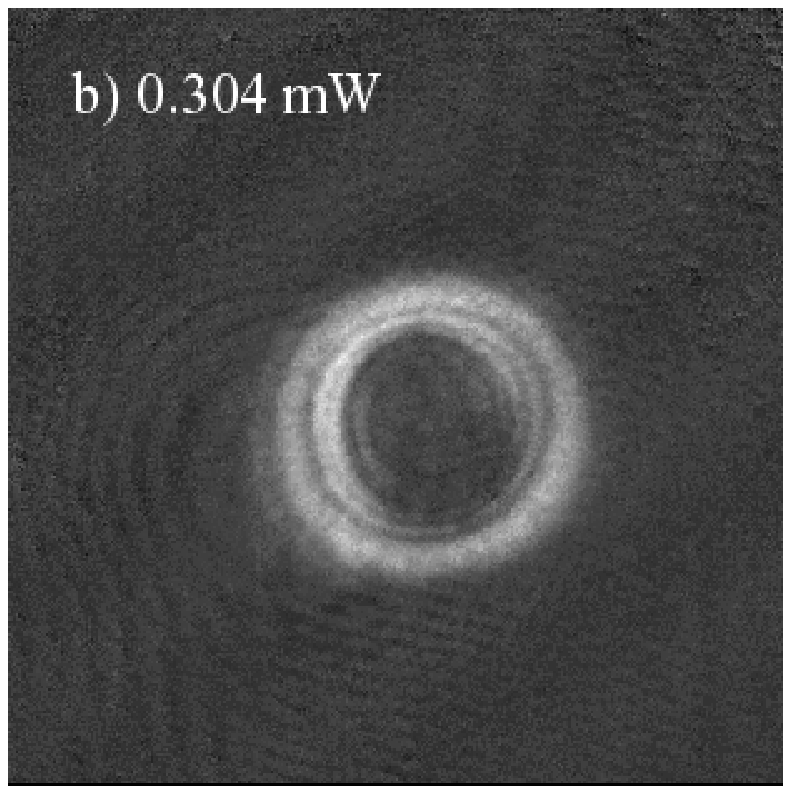}
    \end{minipage}
    \\
    \begin{minipage}{42mm}
      \includegraphics[width=42mm]{
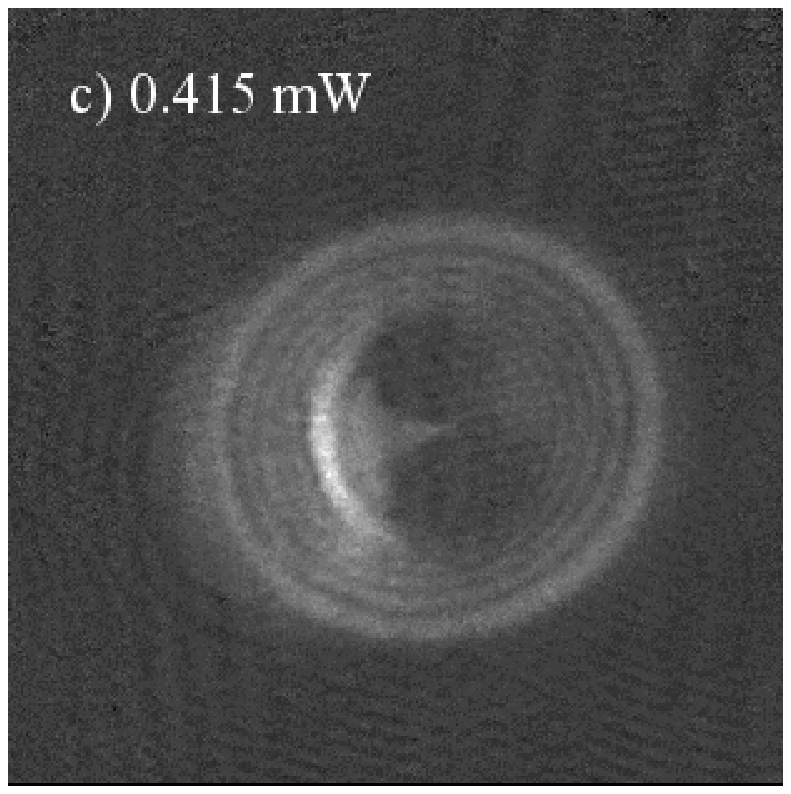}      
    \end{minipage}
    &
    \begin{minipage}{42mm}
      \includegraphics[width=42mm]{
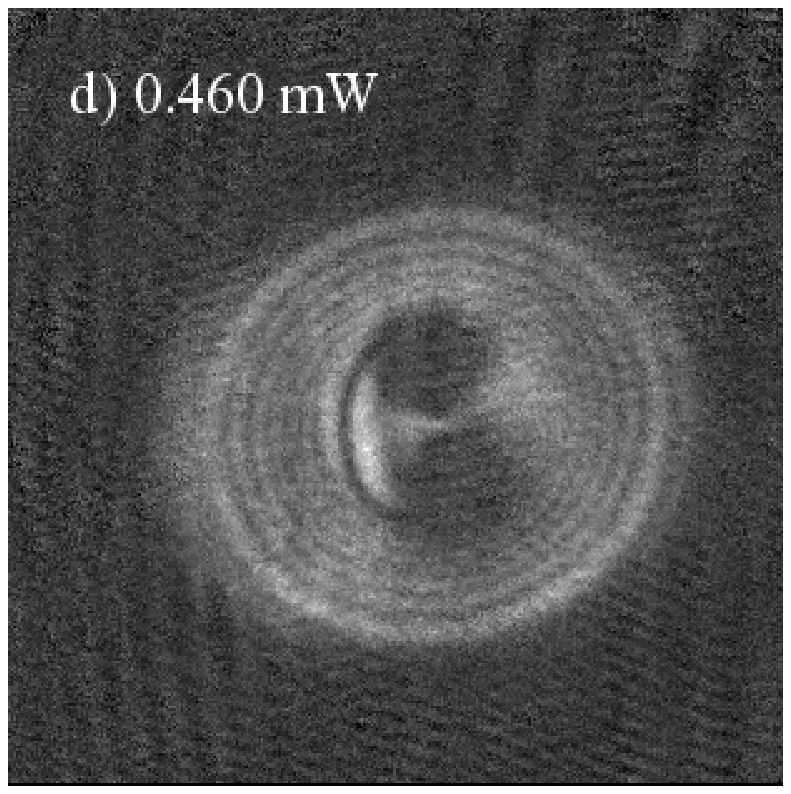}      
    \end{minipage}
    \\
    \begin{minipage}{42mm}
      \includegraphics[width=42mm]{
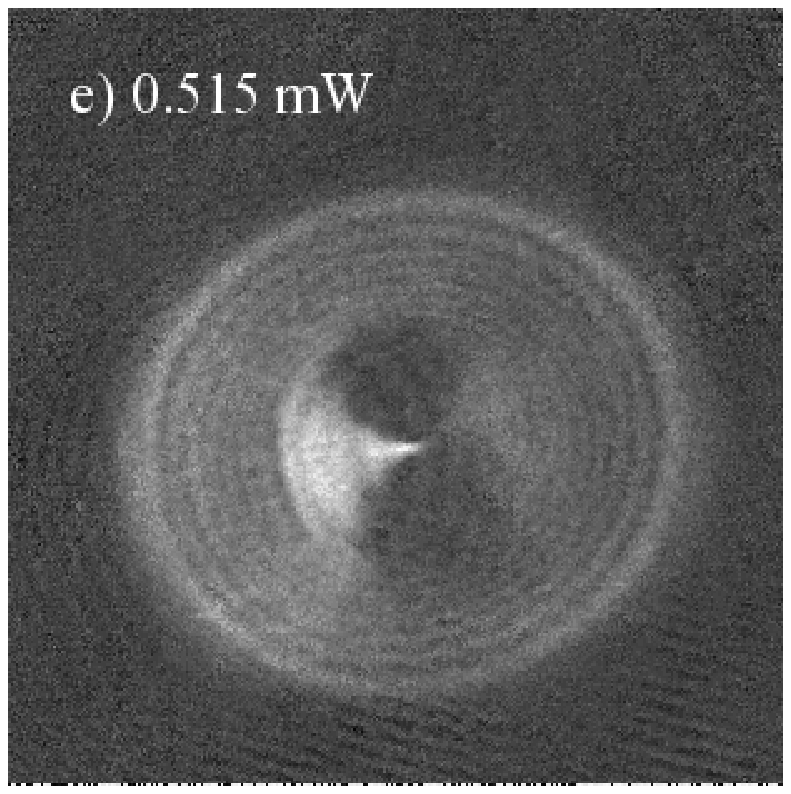}      
    \end{minipage}
    &
    \begin{minipage}{42mm}
      \includegraphics[width=42mm]{
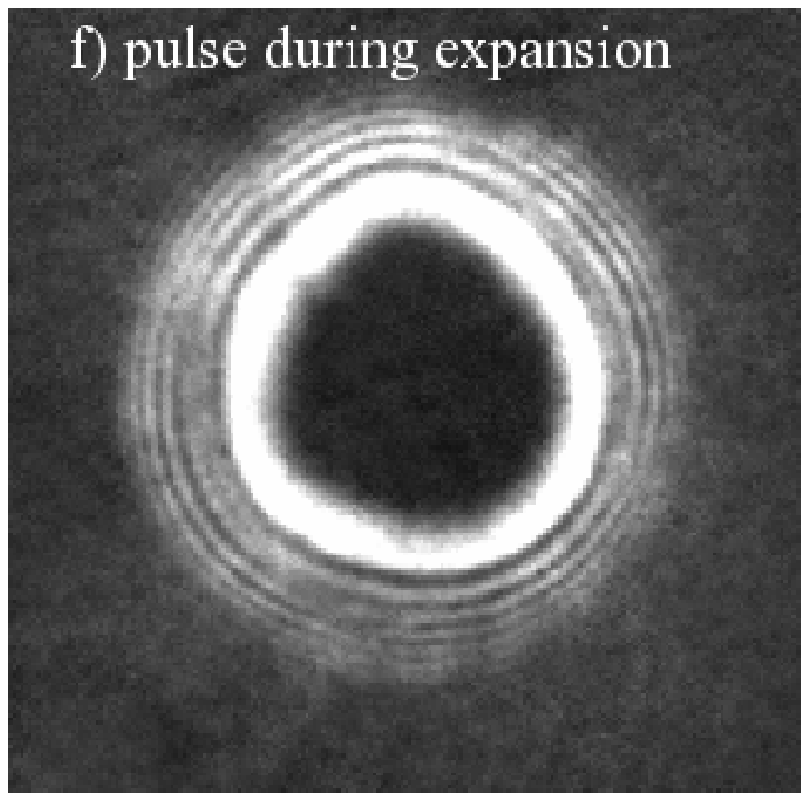}      
    \end{minipage}
  \end{tabular}
  \caption{Absorption images of blast pulse experiments with a BEC.
    a-e) Pulse applied before expansion.  f) Pulse applied during
    expansion.}
  \label{fig:shock_pics1}
\end{figure}
A sequence of five images taken at the end of experimental runs with
different laser pulse intensities is shown in
Fig. \ref{fig:shock_pics1}(a-e).  For this sequence, a 5 ms long pulse
was sent through the BEC center directly before the start of a 50 ms
long anti-trapped expansion.  The laser waist was 13.5 microns.  For
comparison, the diameter of the BEC in the radial direction was
approximately 65 microns.  The laser power is given in the images.
All images were taken at the end of the anti-trapped expansion along
the $z$-axis, which is also the direction of the blast pulse.  For
weak blast pulse intensities (Fig.  \ref{fig:shock_pics1}a,b),
essentially one broad ring of high density is seen, which is due to
the fact that the laser pulse has pushed atoms radially outwards.
When the blast pulse intensity is increased, a system of many
concentric rings appears (\ref{fig:shock_pics1}c-e).

The outcome of a second type of experiment is shown in
Fig. \ref{fig:shock_pics1}f.  By pulsing the blast laser during
(instead of before) the anti-trapped expansion, we can image a
situation where the compressional ring has not run through the
condensate yet.  For this image, a 5 ms long pulse with a power of 1.9
mW and a beam waist of 20 microns was used, starting 9.2 ms after the
beginning of a 55 ms long anti-trapped expansion.  In this case, an
oscillatory wave structure is seen on the outside of the compressional
ring.  The analytical discussion together with the numerical studies
presented in this paper reveal that for both experiments the
oscillatory wave structure is a direct consequence of dispersive shock
waves which are fundamentally different from classical shock waves.

Finally, we note that the peculiar wedge shaped appearance of the
central BEC region in Fig. \ref{fig:shock_pics1}(c-e) is due to a
slight, unavoidable deviation of the laser beam shape from cylindrical
symmetry.  This asymmetry also leads to the slightly elliptical
appearance of the whole BEC in these images.

\begin{figure}
  \centering
  \begin{tabular}{cc}
    \begin{minipage}{42mm}
      \includegraphics[width=42mm]{
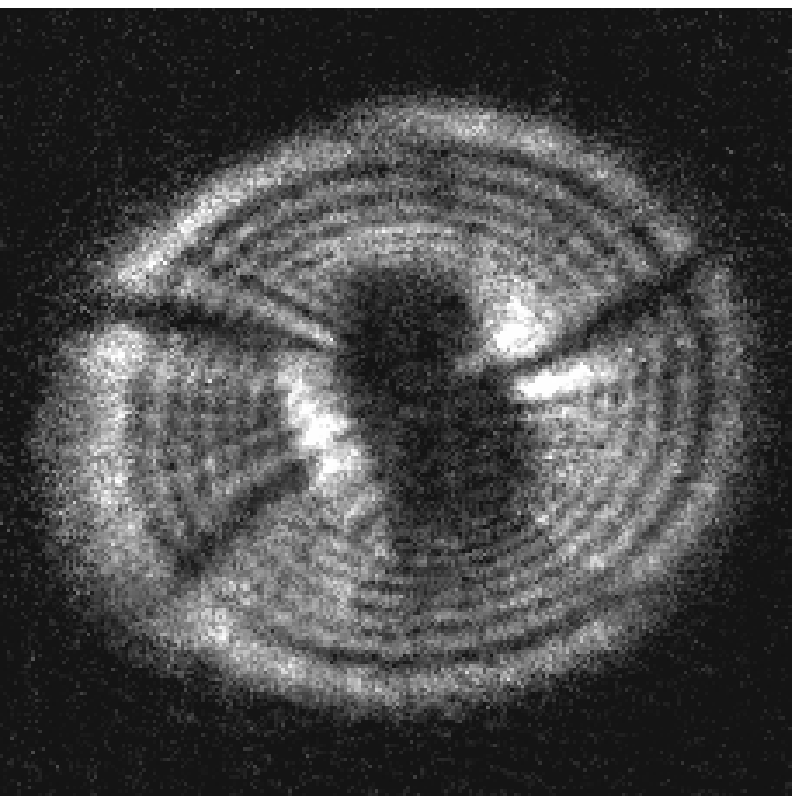}
    \end{minipage}
    &
    \begin{minipage}{42mm}
      \includegraphics[width=42mm]{
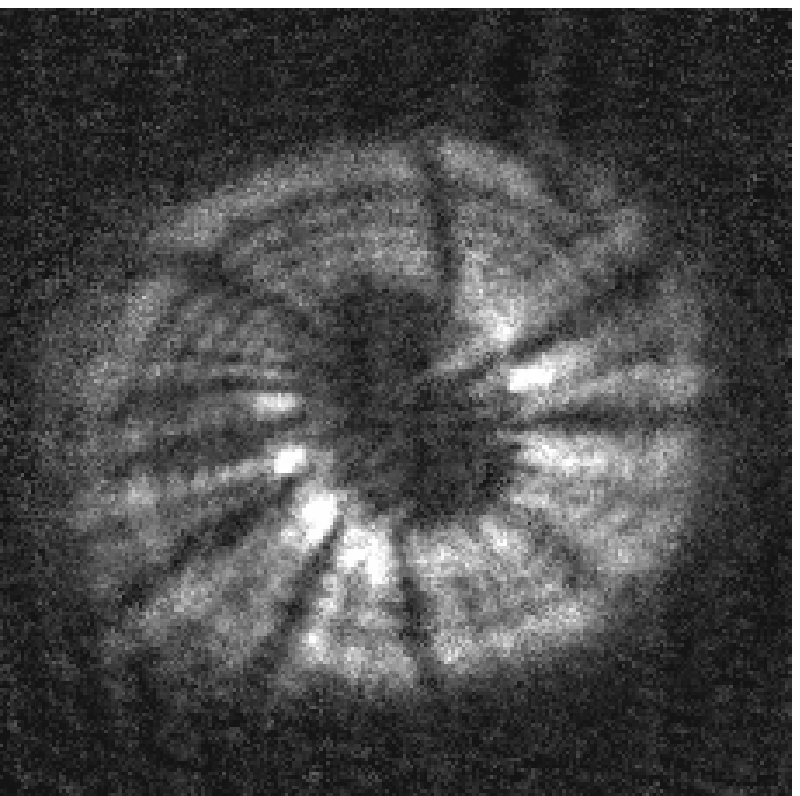}
    \end{minipage}
  \end{tabular}  
  \caption{Two examples of experiments with blast pulses in slowly
    rotating BECs.}
  \label{fig:rotating}
\end{figure}
By using rotating instead of static condensates, we can also observe
an intriguing alteration of the blast wave pattern.  Blast wave images
in slowly rotating BECs are shown in Fig. \ref{fig:rotating}.  Upon
slow rotation, dark radially directed spokes appear in the condensate,
cutting through the ring shaped pattern familiar from the non-rotating
case.  The number of these spokes increases with increasing rotation
rate, so it is suggestive to attribute them to the presence of
vortices in the rotating BEC.  We speculate that the spokes come about
when vortices are present in the compressional ring formed by the
blast.  As this ring expands and forms the concentric ring system, the
density depressions of the vortices are not filled in due to the
predominantly radial expansion of the compressional ring.  In rapidly
rotating BECs, the presence of strong Coriolis forces lead to a rather
different appearance of blast waves.  The rapidly rotating situation
was discussed in \cite{Engels05}.

\section{Simulations}
\label{sec:numerical-results}
We have performed direct numerical simulations of the GP equation
\eqref{eq:1}, modeling the two types of experiments without rotation
explained in the previous section.

The two experiments depend on when the laser is pulsed. We refer to
these cases as either in trap (\textit{it}) or out of trap
(\textit{ot}) and model them by the following time varying potentials
respectively
\begin{align}
  \label{eq:64}
  V_{it}(r,z,t) &= \left \{
    \begin{array}{cc}
      \tfrac{1}{2}(r^2 + \alpha_z z^2) 
      & t < 0 \\[2mm]
      \tfrac{1}{2}(r^2 + \alpha_z z^2) +
      \frac{P_{it}}{d_{it}^2} e^{-r^2/d_{it}^2} & 0 \le t \le \delta t \\[2mm]
      \tfrac{1}{2}(-\alpha_{r}r^2 +
      \alpha_z z^2) &  \delta t < t \\[2mm] 
    \end{array} \right . \\
  \label{eq:66}
  V_{ot}(r,z,t) &= \left \{
    \begin{array}{cc}
      \tfrac{1}{2}(r^2 + \alpha_z z^2) & 
      t < 0 \\[2mm]
      \tfrac{1}{2}(-\alpha_{r}r^2 +
      \alpha_z z^2) & 0 \le t \le t_* \\[2mm] 
      \begin{array}{c}
        \tfrac{1}{2}(-\alpha_{r}r^2 +
        \alpha_z z^2) \\[1mm]
        + \frac{P_{ot}}{d_{ot}^2} e^{-r^2/d_{ot}^2}
      \end{array}
      &
        t_* \le t \le
        t_*+\delta t  
      \\[2mm]
      \tfrac{1}{2}(-\alpha_{r}r^2 +
      \alpha_z z^2) &  t_* + \delta t < t. \\[2mm] 
    \end{array} \right .
\end{align}
The in trap potential \eqref{eq:64} models a steady state solution
held in trap while a Gaussian laser pulse is applied for the time
$\delta t$.  After the laser pulse, a radial anti-trap potential is
applied \cite{Coddington04}.  This models the experiments shown in
Fig. \ref{fig:shock_pics1}(a-e).  For the out of trap potential
\eqref{eq:66} (modeling Fig. \ref{fig:shock_pics1}f), a steady state
solution is expanded radially.  At the time $t_*$, a Gaussian laser
pulse is applied with duration $\delta t$ followed by continued radial
expansion.  The Gaussian laser has width $d$ and intensity
proportional to $P$.  The out of trap potential has the effect of
generating an outward, radial velocity in the BEC before the laser
pulse is applied whereas the in trap potential does not.

Modeling the non-rotating experiments presented in the previous
section gives the parameter values: $t_* = 0.48$ (9.2 ms), $\delta t =
0.26$ (5 ms), $\alpha_r = 0.71$ (frequency of inverted harmonic
potential used for expansion $2\pi\cdot 7$ Hz), and $\alpha_z = 0.57$.

The steady state solution is approximated well by the Thomas-Fermi
wavefunction \cite{PeSm02}, a balance between the harmonic trapping
potential and nonlinearity.  However, its use numerically gives rise
to unphysical oscillations.  Therefore, we used an iterative technique
similar to the technique discussed in \cite{Musslima04} to find the
precise, smooth 3D solution of the GP equation.  We provide a brief
outline of the method here.

Assume a stationary solution of the form
\begin{equation}
  \label{eq:6}
  \Psi(\vec{x},t) = e^{-i\mu t/\eps} \phi(\vec{x})
\end{equation}
where $\mu$ is the normalized condensate chemical potential.
Inserting the ansatz \eqref{eq:6} into equation \eqref{eq:1} and
taking its Fourier transform (denoted by $\widehat{\ \ \ \ }$) gives
\begin{equation}
  \label{eq:25}
  (\mu - \tfrac{1}{2}\eps^2 |\vec{k}|^2) \widehat{\phi} = 
  \widehat{V_0 \phi} + \widehat{|\phi|^2 \phi} \equiv \widehat{F}(\phi),
\end{equation}
where $\vec{k}$ is the Fourier wave vector.  Equation \eqref{eq:25}
suggests the iteration
\begin{equation*}
  \widehat{\phi}_{n+1} = \frac{\widehat{F}(\phi_n)}  {\mu -
    \tfrac{1}{2}\eps^2 |\vec{k}|^2}. 
\end{equation*}
For $\mu > 0$, the denominator is singular when $|\vec{k}|^2 =
2\mu/\eps^2$.  Also, the normalization \eqref{eq:16} is not conserved
by this scheme.  Therefore, we introduce the modified iteration scheme
\begin{equation}
  \label{eq:45}
  \begin{split}
    \widehat{\phi}_{n+\frac{1}{2}} &= \frac{ \widehat{F}(\phi_n) + c
      \widehat{\phi}_n} {\mu -
      \tfrac{1}{2}\eps^2 |\vec{k}|^2 + c}, \\
    \widehat{\phi}_{n+1} &=
    \frac{\widehat{\phi}_{n+\frac{1}{2}}}{\sqrt{ \int_{\real^3}
        |\widehat{\phi}_{n+\frac{1}{2}}|^2 \, \ud^3 k}}, \\
    \mu_{n+1} &= \int_{\real^3} \left( \frac{\eps^2}{2} |\vec{k}|^2
      \widehat{\phi}_{n+1} + \widehat{F}(\phi_{n+1}) \right)
    (\widehat{\phi}_{n+1})^* \, \ud^3 k .
  \end{split}
\end{equation}
The positive constant $c$ is introduced by adding $c \widehat{\phi}$
to both sides of equation \eqref{eq:25}.  The chemical potential,
$\mu$, is updated along with the mode $\phi$ by integrating equation
\eqref{eq:25}, with $\widehat{\phi} = \widehat{\phi}_{n+1}$,
multiplied by $(\widehat{\phi}_{n+1})^*$.  This scheme preserves the
normalization condition \eqref{eq:16}.  Given an initial guess
$\phi_0$ and a large enough value for $c$ (we took $c = 1.75$), we
find that the scheme \eqref{eq:45} converges to a 3D ground state
wavefunction $\phi(\vec{x})$ for the GP equation.  This is a general,
fast method for finding the 3D BEC ground state for arbitrary
potentials $V_0$.

The condensate is evolved according to a pseudo-spectral Fourier code
with a standard $4^{\text{th}}$ order Runge-Kutta time stepper.
Radial spatial derivatives are approximated by taking the fast Fourier
transform of $\Psi$ in the radial direction $r$ evenly extended
($\Psi(-r,z,t) = \Psi(r,z,t)$) and then multiplying by $ik$ ($k$ is
the Fourier wavenumber).  The result is then inverted using the
inverse fast Fourier transform giving a fast, spectrally accurate
approximation to $\Psi_r$ \cite{Fornberg98}.  The $\Psi_{zz}$ term is
approximated similarly.  The specific, model equation \eqref{eq:1}
assumes cylindrical coordinates with axial symmetry: $\nabla^2 \equiv
\pdd{}{r} + \frac{1}{r} \pd{}{r} + \pdd{}{z}$.  In the simulations, we
used the laser width as an effective fitting parameter.  We found that
by taking the width, $d$, to be 1.5 times larger than its experimental
value in the in trap case, excellent results were obtained.  For the
out of trap case, the laser width was taken to be half its
experimental value, also giving excellent results.

\begin{figure}
  \centering
  \begin{tabular}{cc}
    $\quad|\Psi|^2\qquad\qquad$ & $\qquad\qquad\qquad\quad(\arg \Psi)_r$
  \end{tabular}
  \includegraphics[scale=.44]{
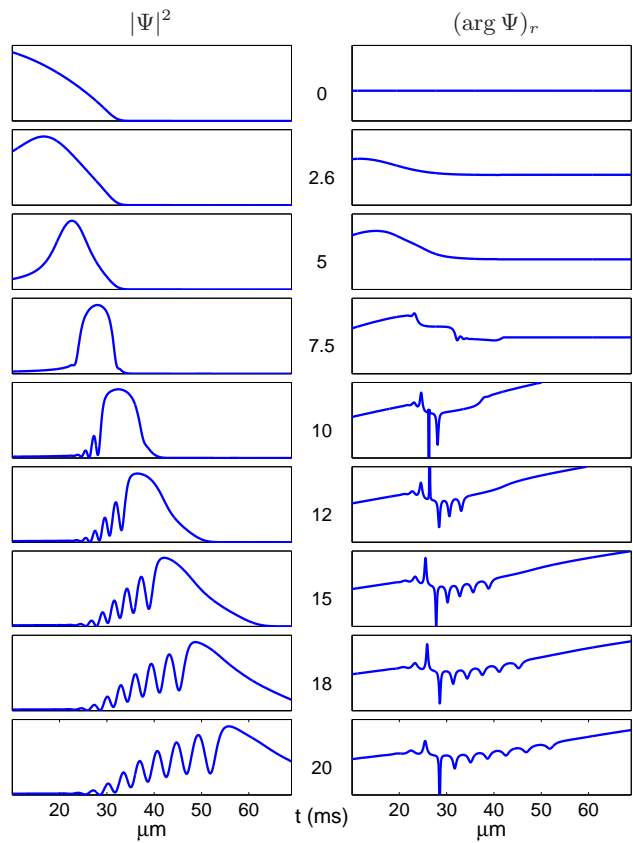}
  \caption{BEC density evolved as in the in trap experiment.  On the
    left is the density $|\Psi(r,0,t)|^2$ and on the right is the
    radial velocity $\pd{}{r} \arg \Psi(r,0,t)$; both are imaged in
    the $z = 0$ plane.  The density scale is not constant throughout
    the sequence; in each frame, the density is scaled to its largest
    value so that the dispersing wave is visible. A high density ring
    forms accompanied by oscillations on its inner side, a DSW.  This
    DSW expands and propagates outward.  Note the time is in the
    center.  All figures depicting a time evolution of the condensate
    in this work follow the same form as this figure.}
  \label{fig:full_experiment_it}
\end{figure}
First we consider the in trap case corresponding to the potential
\eqref{eq:64}.  A plot of the evolution of the condensate as a
function of radial distance is shown in Fig.
\ref{fig:full_experiment_it}.  On the left is the normalized density
as given by the square modulus of the wavefunction plotted in the $z =
0$ plane $|\Psi(r,z=0,t)|^2$. On the right is the phase gradient
$\pd{}{r} \arg \Psi(r,z=0,t)$ or radial velocity in the $z = 0$ plane.
The normalized laser width and power used were $d_{it} = 0.41$ and
$P_{it} = 0.70$ (corresponding to a laser waist of 20 $\mu$m and a
power of 0.46 mW as in Fig. \ref{fig:shock_pics1}d).

\begin{figure}
  \centering
  \begin{tabular}{cc}
    \begin{minipage}{42mm}
      \includegraphics[width=42mm,clip=true,trim=25 20 5 35]{
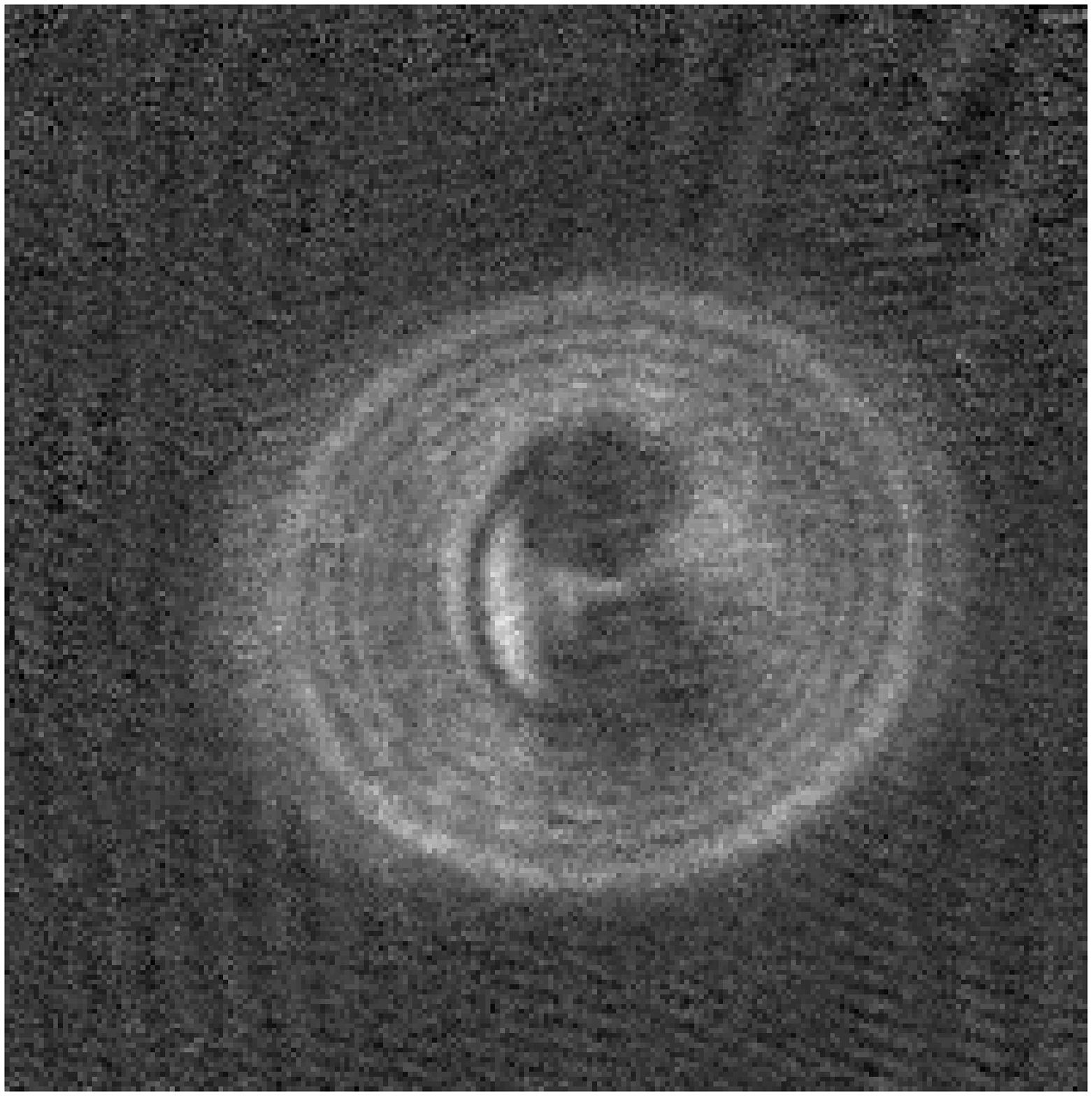}
    \end{minipage}
    &
    \begin{minipage}{42mm}
      \includegraphics[width=42mm]{
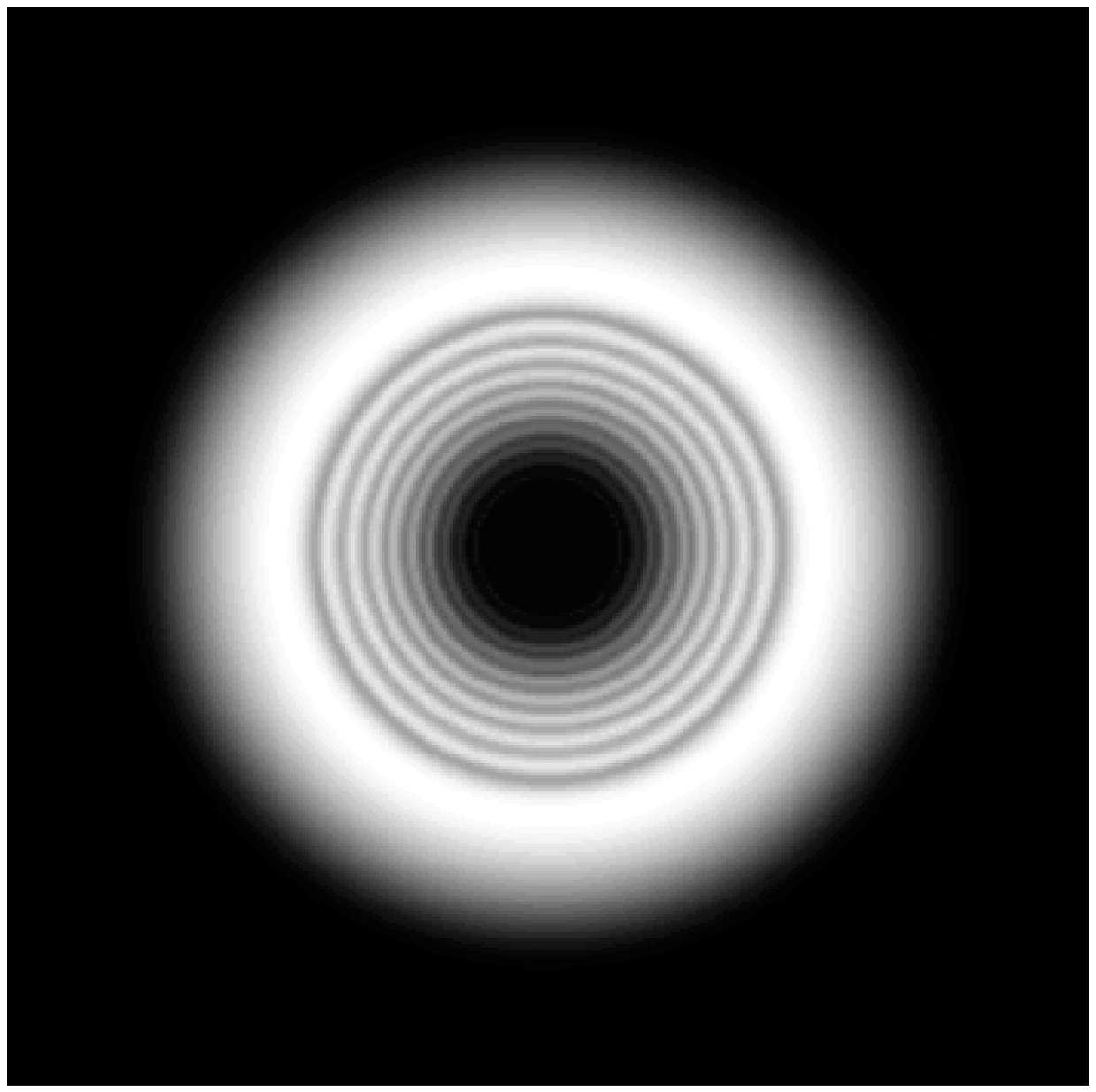}
    \end{minipage}
  \end{tabular}  
  \caption{Comparison of the condensate density from the in trap
    experiment (left) and numerical simulation (right) using the
    potential $V_{it}$, equation \eqref{eq:65}.  The image from
    simulation is a contour plot of the function $\int |\Psi(r,z,t)|^2
    \, \ud z$, modeling the experimental imaging process where the
    photo was taken along the $z$ axis.  Both the simulation and
    experimental pictures were taken at $t = 55$ ms.  The approximate
    diameter of the condensate cloud from simulation is 850 $\mu$m and
    from experiment is 775 $\mu$m.}
  \label{fig:it_compare}
\end{figure}

Fig. \ref{fig:full_experiment_it} describes the following evolution.
The condensate forms a high density ring ($t = 0$ to 5 ms $= \delta t$
in non-dimensional form) due to the applied laser pulse.  When the
ring is steep enough ($t = 7.5$ ms), oscillations develop on the inner
side of the ring ($t = 10$ ms).  This oscillatory region expands
radially due to the inversion of the trap potential to an anti-trap
potential.  The expansion continues until time $t = 55$ ms when an
image is taken to compare with experiment (Fig. \ref{fig:it_compare}).

In Fig. \ref{fig:it_compare}, we show that the numerical simulation
and the in trap experiment presented in the previous section
(Fig. \ref{fig:shock_pics1}d) are in good qualitative agreement.  The
condensate features at $t = 20$ ms in
Fig. \ref{fig:full_experiment_it} expand due to atomic repulsion and
the anti-trap potential giving the contour plot of the density at
$t=55$ ms (corresponding to the end of the experiment) in
Fig. \ref{fig:it_compare}.  The experimental picture
\ref{fig:shock_pics1}d is reproduced in Fig. \ref{fig:it_compare} left
for convenience.  The simulation used all nominal values for
parameters except the laser waist ($d$ in non-dimensional units) which
was taken to be $20\, \mu$m rather than the experimental value of
$13.5\, \mu$m.  We speculate that the difference might be due to the
slight deviation of the experimental beam profile from a perfect
Gaussian, as indicated by the asymmetry of the central regions in
Fig. \ref{fig:shock_pics1}.

\begin{figure}
  \centering
  \begin{tabular}{cc}
    $\quad|\Psi|^2\qquad\qquad$ & $\qquad\qquad\qquad\quad(\arg \Psi)_r$
  \end{tabular}
  \includegraphics[scale=.47]{
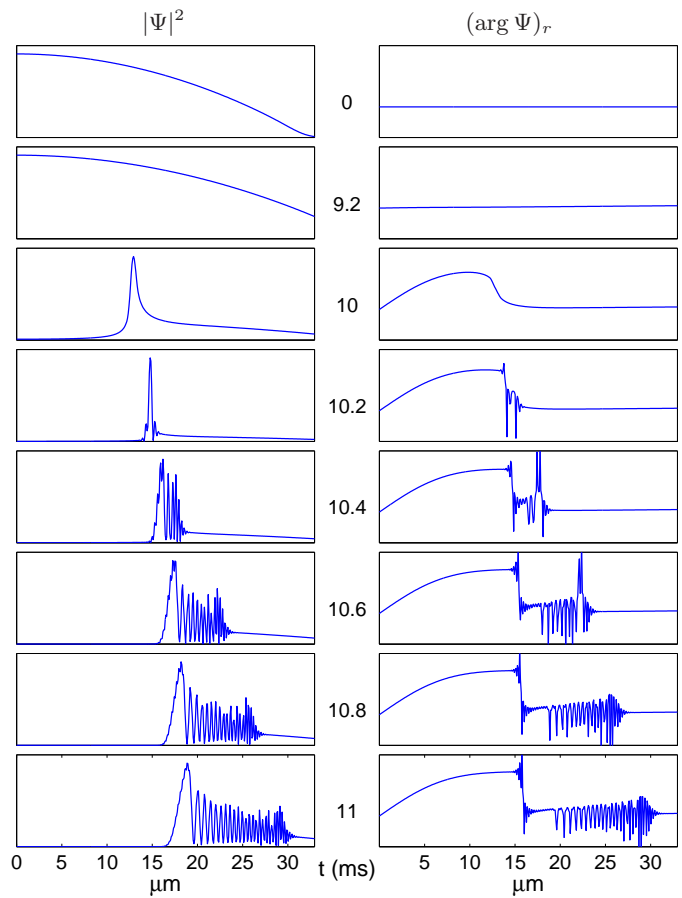}
  \caption{BEC density evolved as in the out of trap experiment.  On
    the left is the density and on the right is the radial velocity,
    both imaged in the $z = 0$ plane.  A high density ring forms
    accompanied by oscillations on its inner and outer sides, two
    DSWs.  These DSWs quickly interact and propagate outward.}
  \label{fig:full_experiment}
\end{figure}
Next we consider the out of trap case corresponding to the potential
$V_{ot}$ \eqref{eq:66}.  A plot of the evolution of the condensate in
the $z = 0$ plane as a function of radial distance $r$ is shown in
Fig. \ref{fig:full_experiment}.  Initially, the condensate expands in
the radial direction due to the anti-trap potential ($t = 0$ to 9.2 ms
or $t = t_*$ in normalized units).  A ring of high density forms while
the laser is on ($t = 10$ ms), similar to the previous in trap case.
When the steepness of this ring is large enough, oscillations start to
develop on the inner \emph{and} outer sides of the ring ($t = 10.2$
ms).  These two oscillatory regions expand, quickly overlapping one
another, giving rise to more complicated \emph{multi-phase} type
behavior with a propagating wave front that continues out radially ($t
\ge 10.4$ ms).  This behavior is due to the initial velocity imparted
to the condensate by the anti-trap potential.  The normalized laser
width and power were $d_{ot} = 0.21$ and $P_{ot} = 2.88$ corresponding
to a laser waist of 10 $\mu$m (half the experimental value of 20
$\mu$m) and a power of 1.9 mW.

\begin{figure}
  \centering
  \begin{tabular}{cc}
    \begin{minipage}{42mm}
      \includegraphics[width=42mm]{
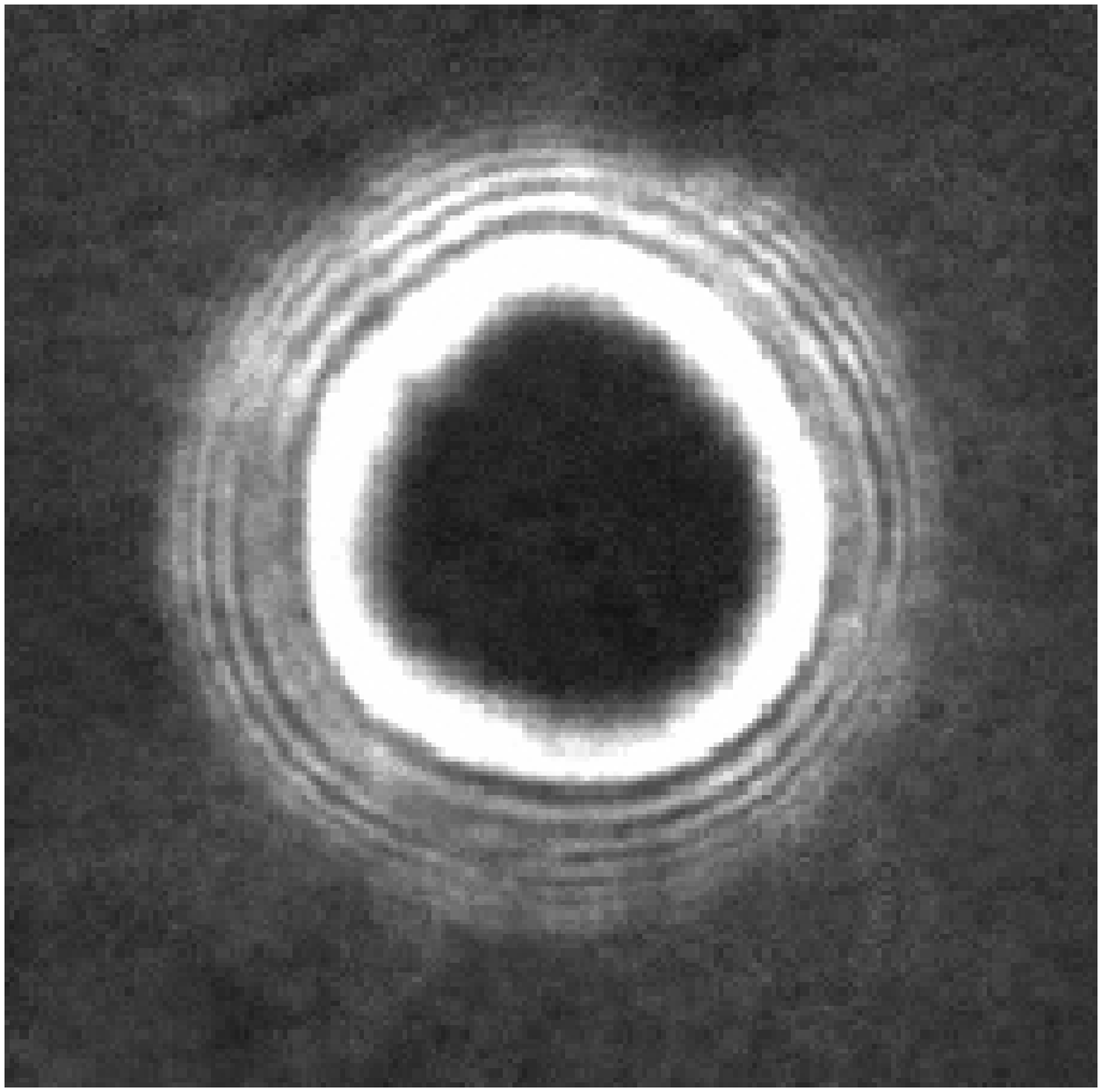}
    \end{minipage}
    &
    \begin{minipage}{42mm}
      \includegraphics[width=42mm,clip=true,trim=10 10 10
      10]{
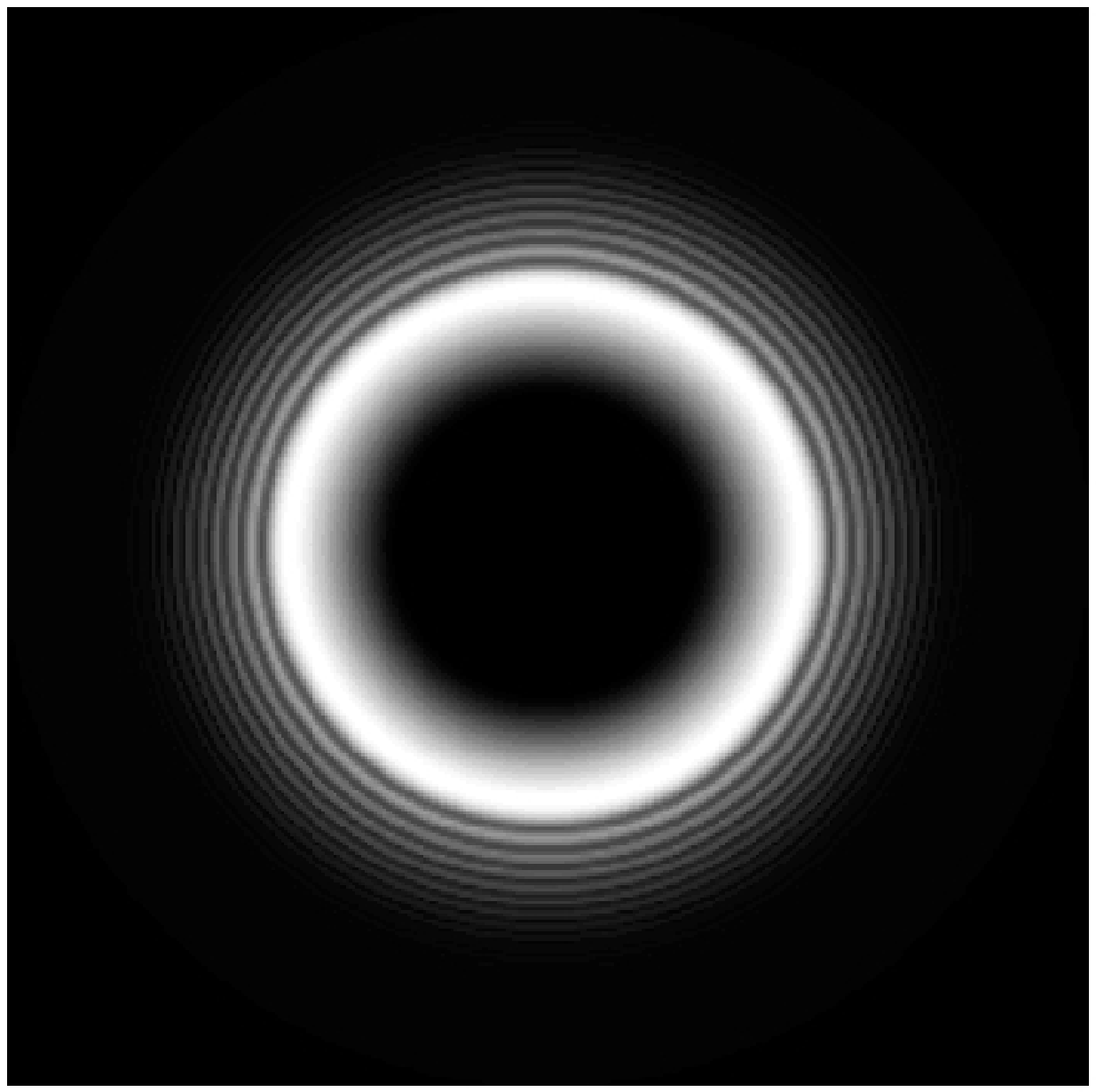}
    \end{minipage}
  \end{tabular}  
  \caption{Comparison of the condensate density from the out of trap
    experiment (left) and numerical simulation using the potential
    $V_{ot}$ \eqref{eq:66} (right).  The image from simulation is a
    contour plot of the function $\int |\Psi(r,z,t)|^2 \, \ud z$,
    modeling the experimental imaging process.  The approximate
    diameter of the condensate cloud from simulation is 116 $\mu$m and
    from experiment is 363 $\mu$m.}
  \label{fig:ot_compare}
\end{figure}

In Fig. \ref{fig:ot_compare}, we show that the numerical simulation of
the out of trap experiment presented in the previous section
(Fig. \ref{fig:shock_pics1}f) show good qualitative agreement.  For
this comparison, the numerically determined condensate density is
shown at the time $t = 18.0$ ms, the result of continued expansion
from the state at the bottom of Fig. \ref{fig:full_experiment}.  The
experimental picture was taken at $t = 55$ ms.

The rest of this paper is concerned with understanding the oscillatory
structures that developed in Figures \ref{fig:full_experiment_it} and
\ref{fig:full_experiment} as we will argue that they are dispersive
shock waves with the oscillations in the latter Figure caused by the
interaction of two DSWs. 

\section{Classical and Dispersive Shock Waves}
\label{sec:class-disp-shock}
In this section, the notions of classical shock waves and dispersive
shock waves are discussed. We provide a theoretical basis for the
experimental and numerical results presented in the previous sections.
Using the classical gas dynamics analogy, it is shown that shocks in a
BEC are fundamentally different from those in the classical case.  The
analytical methods to understand classical shocks and DSWs in one
dimension are presented along with explicit results for DSW speeds and
shock structure.  Finally, detailed numerical investigations in 1D,
2D, and 3D are presented to show that the qualitative behavior of a
DSW in 3D is captured by the 1D case and that the experiments
presented earlier do give rise to DSWs.

We begin by studying the simplest nonlinear dissipative and dispersive
equations that, under a suitable limit, give rise to shock waves, the
Burgers and Korteweg-deVries equations.  This provides a foundation to
understand the differences between dissipative and dispersive shocks.
The theory of classical shock waves is well developed (cf.
\cite{CoFr48,Lax73,LeFloch02}) so we will give only a brief synopsis
suitable for comparison with the much less developed dispersive shock
wave case.

\subsection{Classical Shock Waves, Burgers' Equation}
\label{sec:burgers-equation}
\begin{figure}
  \centering
  \includegraphics[scale=.35]{
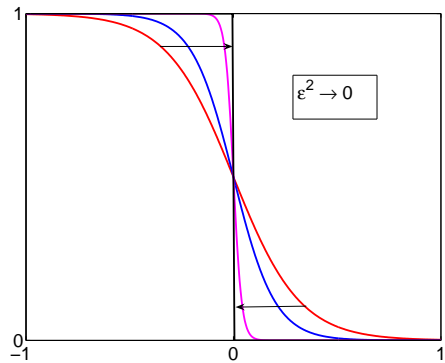}
  \caption{Dissipative Burgers' equation shock solution (\ref{eq:54}) with
    $\eps^2 \to 0$, converging to a traveling discontinuity or
    classical shock wave.}
  \label{fig:classical_shock}
\end{figure}

The classical dissipative shock in one dimension is modeled by
Burgers' equation
\begin{equation}
  \label{eq:10}
    u_t + (\tfrac{1}{2}u^2)_x = \eps^2 u_{xx},
\end{equation}
where $\eps^2$ is a measure of the dissipation and $u$ represents, for
example, a density.  Equation (\ref{eq:10}) admits traveling wave
solutions with a hyperbolic tangent profile (see Fig.
\ref{fig:classical_shock})
\begin{equation}
  \label{eq:54}
  u(x,t;\eps) = \tfrac{1}{2} +
  \tfrac{1}{2}\tanh\left\{-\tfrac{1}{4\eps^2}(x-\tfrac{1}{2}t)\right\} . 
\end{equation}
The speed of this wave is $v_c = 1/2$, independent of $\eps$.  In the
limit $\eps \to 0$, the (smooth) tanh profile converges pointwise to
the discontinuous function
\begin{equation}
  \label{eq:53}
  \lim_{\eps \to 0} u(x,t;\eps) = 
  u(x,t) = u(x/t) = \left\{
    \begin{array}{lc}
      1 & x/t < \tfrac{1}{2} \\[2mm]
      0 & x/t > \tfrac{1}{2}
    \end{array}
  \right. .
\end{equation}
The above formula is a mathematical description of a classical shock
wave.  The limiting process $\eps \to 0$, $\eps \neq 0$, in equation
(\ref{eq:10}) is a \emph{dissipative regularization} of the
conservation law
\begin{equation}
  \label{eq:55}
  u_t + (\tfrac{1}{2}u^2)_x = 0. 
\end{equation}

The initial value problem (IVP) for equation \eqref{eq:55} with the
initial data 
\begin{equation}
  \label{eq:119}
  u(x,0) = \left \{
    \begin{array}{cc}
      1 & x<0 \\
      0 & x>0
    \end{array} \right.,
\end{equation}
is not well posed because the spatial derivative $u_x$ is undefined at
the origin.  To see this, note that equation \eqref{eq:55} shows that a
wave $u(x,t)$ propagates with a speed equal to the value of $u$ at
that point.  Initially, for $x < 0$, the speed is $1$ whereas for
$x>0$, the speed is 0 so $u$ will \emph{break} or become multi-valued
at the origin for any $t >0$.  The classical theory of shock waves
remedies this problem by considering \emph{weak solutions} and
invoking a \emph{jump} or Rankine-Hugoniot condition at a
discontinuity which relates the value of $u$ on the left ($u_l$) and
the right ($u_r$) to the speed $v$ of the shock wave.  A weak solution
$u(x,t)$ for the conservation law \eqref{eq:55} satisfies the integral
formulation
\begin{equation}
  \label{eq:13}
  \frac{d}{dt} \int_a^b u(x,t) \, \ud x + \tfrac{1}{2}
  ( u(b,t)^2 - u(a,t)^2) = 0,
\end{equation}
for any $a$, $b$ such that $-\infty < a < b < \infty$, thus allowing
discontinuities in $u(x,t)$ as a function of $x$.  The jump condition
for Burgers' equation, derived from \eqref{eq:13} assuming a uniformly
traveling discontinuity with values $u_l$ and $u_r$ to the left and
right of the discontinuity respectively, is $v = (u_l+u_r)/2$ or $v =
\frac{1}{2}$ for the initial data \eqref{eq:119}, which is exactly the
speed of the Burgers shock \eqref{eq:54}.  This simple example
suggests that finding the dissipative regularization of \eqref{eq:55}
(equation \eqref{eq:10} with $\eps^2 \to 0$) is equivalent to solving
the conservation law \eqref{eq:55} with the jump condition $v = (u_l +
u_r)/2$ at each discontinuity.  Indeed, this is generally true,
assuming the entropy condition $u_l > u_r$ is satisfied \cite{Lax73}.

 When the entropy condition at a discontinuity is not satisfied, $u_l <
u_r$, a shock wave solution is not appropriate.  The correct choice is
a rarefaction wave which is continuous for $t > 0$.  Assuming that the
solution depends on the self-similar variable $\xi = x/t$, equation
\eqref{eq:55} becomes
\begin{equation}
  \label{eq:65}
  u'(u - \xi) = 0.
\end{equation}
Solutions to \eqref{eq:65} are constants or $u(x,t) = x/t$, the latter
corresponding to a rarefaction wave.  Then the weak solution for
initial data $u(x,0) = 0,~x<0$, $u(x,0) = 1, ~x>0$ is
\begin{equation}
  \label{eq:5}
  u(x,t) = u(x/t) = \left \{
    \begin{array}{cc}
      0 & x/t < v^- \\[2mm]
      \frac{x}{t} & v^- < x/t < v^+ \\[2mm]
      1 & v^+ < x/t
    \end{array}
  \right. .
\end{equation}
This rarefaction wave has two associated speeds $v^-$ and $v^+$: $v^-
= 0$ at the interface between the constant left state $u = 0$ and the
self-similar solution $u = x/t$ and $v^+ = 1$ at the interface between
the constant right state $u = 1$ and the self-similar solution $u =
x/t$.

The general case of a system of conservation laws is written
\begin{equation*}
  \vec{u}_t + (\vec{F}(\vec{u}))_x = 0,
\end{equation*}
where $\vec{u}$ is a vector ``density'' and $\vec{F}$ is the vector
``flux''.  The IVP for this $n$-dimensional system with constant step
initial data $\vec{u}(x,0) = \vec{u}_l$, $x < 0$ and $\vec{u}(x,0) =
\vec{u}_r$, $x > 0$, assuming a dissipative regularization, is called
a Riemann problem (note that the jump is specified at $x = 0$).  The
jump condition at a shock with speed $v$ is
\begin{equation}
  \label{eq:63}
  v (\vec{u}_r - \vec{u}_l) = \vec{F}(\vec{u}_r) - \vec{F}(\vec{u}_l).
\end{equation}
It is well known (see e.g. \cite{Lax73}) that the solution to the
Riemann problem, assuming certain properties of the flux $\vec{F}$, is
self-similar consisting of $n+1$ constant states ``connected'' by
shock waves or rarefaction waves.  That is
\begin{equation}
  \label{eq:22}
  \vec{u}(x/t) = \left \{
    \begin{array}{cc}
      \vec{u}_0 & x/t < v_1^- \\[1mm]
      \vec{w}_1(x/t) & v_1^-  < x/t < v_1^+  \\[1mm]
      \vec{u}_1 & v_1^+ < x/t < v_2^- \\
      \vdots & \vdots \\
      \vec{w}_n(x/t) & v_n^- < x/t < v_n^+ \\[1mm]
      \vec{u}_n & v_n^+ < x/t 
    \end{array}
  \right .,
\end{equation}
where each $\vec{u}_i$ is constant and $\vec{w}_i(x/t)$ represents a
shock or rarefaction wave solution.   The Lax entropy condition
necessary for the existence of the shock wave $\vec{w}_i$ is
\cite{Le02}
\begin{equation}
  \label{eq:60}
  \begin{split}
    \lambda_i(\vec{u}_{i-1}) &> v_i^- = v_i^+ \equiv v_i >
    \lambda_i(\vec{u}_i),
    \\
    \lambda_k(\vec{u}_{i-1}) &< v_i \quad \text{and} \quad
    \lambda_k(\vec{u}_i) < v_i \quad k < i, \\
    \lambda_k(\vec{u}_{i-1}) &> v_i \quad \text{and} \quad
    \lambda_k(\vec{u}_i) > v_i \quad k > i.
  \end{split}
\end{equation}
A shock wave $\vec{w}_i$ has one speed of propagation so the
$2i^{\text{th}}$ inequality in \eqref{eq:22} is replaced by $x=v_it$.
The $\lambda_i$ in \eqref{eq:60} are the eigenvalues of the matrix
with entries at $(i,j)$
\begin{equation*}
  \left( \pd{F_i(\vec{u})}{u_j} \right),
\end{equation*}
numbered so that $\lambda_1 < \lambda_2 < \cdots < \lambda_n$.  In
addition to the entropy condition for the $i^{\text{th}}$ wave
$\vec{w}_i$ to shock, the jump condition \eqref{eq:63} is satisfied
for $v = v_i$, $\vec{u}_l = \vec{u}_{i-1}$, and $\vec{u}_r =
\vec{u}_i$.  Whereas a shock has just one speed, associated with every
rarefaction wave solution $\vec{w}_i(x/t)$ are two speeds $v_i^-$
and $v_i^+$.

The established theory of classical shock waves involves dissipative
regularizations of conservation laws.  For initial step data, the
Riemann problem, there are two types of self-similar solutions of
interest, a shock wave and a rarefaction wave.  In the next section,
we study what happens in the region nearby breaking when a dispersive,
rather than dissipative, term is used to regularize the conservation
law.  It will be shown that self-similarity plays a crucial role and
that a dispersive shock wave corresponds, in some sense, to a simple
rarefaction wave solution of a system of conservation laws.

\subsection{Dispersive Shock Waves, Korteweg-deVries Equation}
\label{sec:kort-devr-equat}
As a simple model of dispersive shock waves (DSWs) (e.g. in plasmas
\cite{GaMo60}), we consider the Korteweg-deVries (KdV) equation
\begin{equation}
  \label{eq:11}
  u_t + (\tfrac{1}{2}u^2)_x = -\eps^2 u_{xxx},
\end{equation}
for $\eps^2 \ll 1$.  We investigate the behavior of the solution to
the initial value problem (IVP)
\begin{equation}
\label{eq:14}
  u(x,0;\eps) = \left \{
    \begin{array}{cc}
      1 & x\le0 \\
      0 & x>0
    \end{array}
    \right.
\end{equation}
as $\eps^2 \to 0$. This is a Riemann problem in the context of a
dispersive regularization of the conservation law (\ref{eq:55}) with
no inherent dissipation.

\begin{figure}
  \centering
  \includegraphics[scale=.42]{
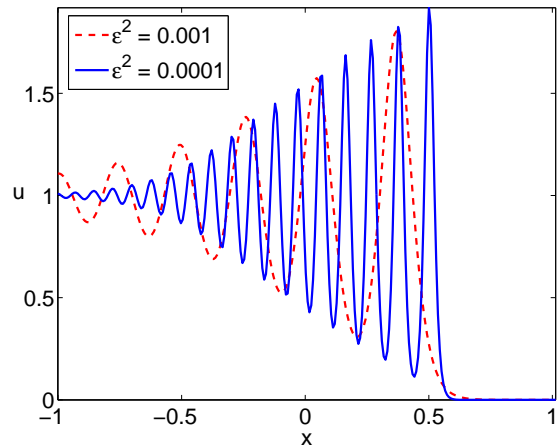}
  \caption{Numerical solutions of equation \eqref{eq:11} for initial
    data \eqref{eq:14} and $\eps^2 = 0.001$ (dashed), $\eps^2 = 0.0001$
    (solid).  As $\eps$ decreases, the wavelength of the oscillations
    decreases.}
  \label{fig:dsw_numerical}
\end{figure}
Figure \ref{fig:dsw_numerical} depicts two numerical solutions to the
IVP \eqref{eq:11} and \eqref{eq:14} for small $\eps^2$.  Oscillations
develop about the initial discontinuity with wavelength proportional
to $\eps$.  Then, as $\eps^2 \to 0$, an infinite number of
oscillations develop. To understand this behavior, one can employ
Whitham's method \cite{Whitham65} which is an extension of the
Krylov/Bogoliubov method of averaging for ODEs to PDEs.  The essence
of the technique is to assume that the KdV equation \eqref{eq:11} has
a uniformly traveling wave solution and average the PDE's conservation
laws over fast oscillations allowing certain parameters (such as
amplitude, wavelength, and speed) to vary slowly in time and space.
Gurevich and Pitaevskii \cite{GuPi} applied Whitham's method to the
IVP considered here with boundary matching where one derives boundary
conditions for the oscillatory region based on continuity of the
averaged flow.  We will follow Gurevich and Pitaevskii's work,
applying the method of initial data regularization presented in
\cite{Kodama99} rather than boundary matching to asymptotically solve
the initial value problem \eqref{eq:11} and \eqref{eq:14} in the limit
$\eps^2 \to 0$.  This limit, denoted $\overline{u}$, is a \emph{weak
  limit} where one averages over the oscillations and is different
from the Burgers' shock strong limit depicted in
Fig. \ref{fig:classical_shock}.  The Whitham method describes the
asymptotic (large $t$) behavior of the slowly modulated oscillatory
region seen in Fig. \ref{fig:dsw_numerical}, which is a dispersive
shock wave, by enabling an explicit calculation of the weak limit
$\overline{u}$.

Just as a discontinuity represents an idealized dissipative shock wave
in a compressible fluid, the weak limit $\overline{u}$ represents an
idealized dispersive shock wave.  Any compressible gas will have a
small but non-zero amount of dissipation.  Since a strong limit
exists, the transition from small to zero dissipation is smooth.  In
the case of a DSW, where a weak limit prevails, the transition from
small to zero dispersion is accompanied by an infinite number of
oscillations.  Thus, any physical DSW with small but non-zero
dispersion will consist of a finite number of oscillations. However,
the weak limit $\overline{u}$ provides an understanding of the
physical DSW structure and its associated speeds.  As we will show, it
also enables clear comparisons between classical and dispersive shock
waves.

\begin{figure}
  \centering
  \includegraphics[scale=.44]{
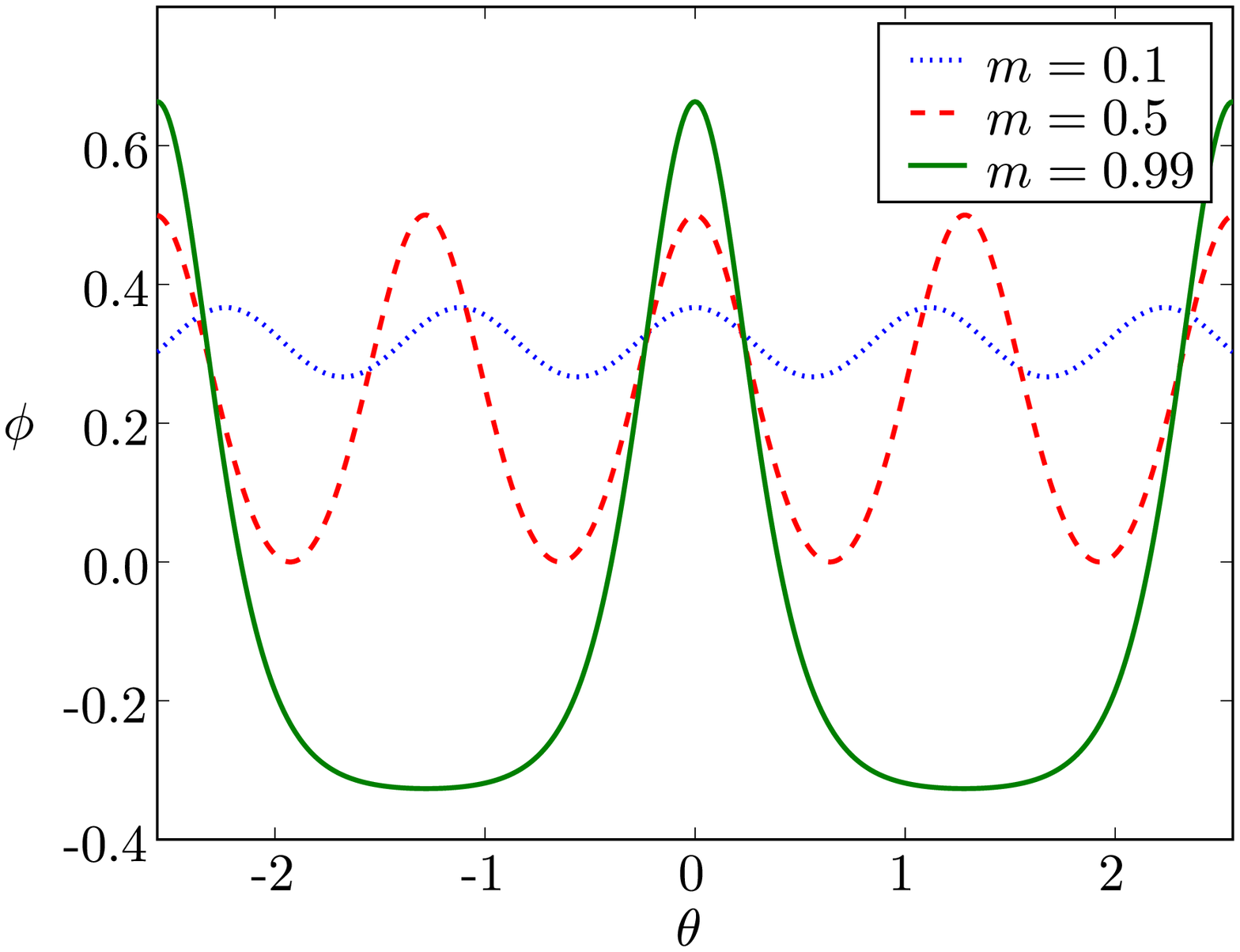}
  \caption{Elliptic function solutions to the KdV equation for several
    choices of the parameters $\{r_i\}$.  The solution converges to a
    constant as $m \to 0$ and to the soliton $\sech$ profile as $m \to
    1$.}
  \label{fig:elliptic}
\end{figure}

The first step in the Whitham averaging method is to obtain a
quasi-stationary periodic solution.  Assuming the traveling wave
ansatz, $u(x,t;\eps) = \phi(\theta)$, $\theta = (x-Vt)/\eps$, equation
\eqref{eq:11} reduces to
\begin{equation*}
  -V \phi' + \phi \phi' + \phi''' = 0. 
\end{equation*}
Integrating this ODE twice we obtain
\begin{equation*}
  (\phi')^2 = -\frac{1}{3}(\phi^3 - 3V\phi^2
  + A \phi + B) \equiv \tfrac{1}{3} P(\phi) ~,
\end{equation*}
with $A$ and $B$ arbitrary integration constants.  Solutions to
equations of this form, when $P(\phi)$ is a cubic or quartic
polynomial, are elliptic functions.  We write the polynomial $P$ in
terms of its roots
\begin{equation*}
  P(\phi) = (\lambda_1 - \phi)(\lambda_2 - \phi) (\lambda_3 - \phi),
  \quad \lambda_1 \le \lambda_2 \le \lambda_3. 
\end{equation*}
For convenience, we make the following linear transformation
\begin{gather*}
  r_1 = \tfrac{1}{2}(\lambda_1 + \lambda_2), ~
  r_2 = \tfrac{1}{2}(\lambda_1 + \lambda_3), ~
  r_3 = \tfrac{1}{2}(\lambda_2 + \lambda_3), \\
  r_1 \le r_2 \le r_3.
\end{gather*}
Then the elliptic function solution can be written as (see
\cite{ByrdFriedman54})
\begin{equation}
  \label{eq:28}
  \begin{split}
    \phi(\theta) &=\, r_1+r_2-r_3 + 2(r_3 - r_1)\,\text{dn}^2 \! \left(
      \sqrt{\frac{r_3-r_1}{6}}
      \theta ;m \right) \\
    m &= \frac{r_2 -r_1}{r_3 - r_1}, ~ \theta = \frac{x - Vt}{\eps}, ~ V
    = \tfrac{1}{3}(r_1 + r_2 + r_3).
  \end{split}
\end{equation}
This is an exact solution to equation \eqref{eq:11} with three free
parameters $\{r_i\}$ related to the amplitude: $\max(\phi)-\min(\phi) =
2(r_2-r_1)$, speed $V$, and wavelength
\begin{equation*}
  L = 2 K(m) \sqrt{\frac{6}{r_3-r_1}},
\end{equation*}
where $K(m)$ is the complete elliptic integral of the first kind.
Note that $L$ is obtained from the periodicity of the dn function
\eqref{eq:28} i.e. $[\sqrt{\frac{r_3-r_1}{6}} \theta] = 2K(m)$ where
$[\cdot]$ is the period of the argument.  The parameter $m$ is the
modulus of the elliptic function.  See Fig.  \ref{fig:elliptic} for a
plot of $\phi$ for various values of $m$.  There are two limiting
behaviors $\text{dn}(y;0) = 1$ and $\text{dn}(y;1) = \sech(y)$, the
solitary wave solution.

The basic idea behind Whitham theory is in the process of averaging
over ``fast'' oscillations.  This yields the behavior of the weak
limit, $\overline{u}$, of equation (\ref{eq:11}) as $\eps \to 0$.
Since $\eps$ is assumed to be much smaller than 1 in
eq. \eqref{eq:11}, the phase $\theta = (x-Vt)/\eps$ is a fast
variable.  We assume that modulations of this periodic solution take
place on the scale of the ``slow'' variables $x$ and $t$.  Then the
average of $\phi$ is
\begin{equation}
  \label{eq:26}
  \begin{split}
    \overline{\phi}(x,t) &= \frac{1}{L} \int_0^{L} \phi(\theta,x,t) \,
    \ud \theta \\
    &= r_1(x,t) + r_2(x,t) - r_3(x,t) + \\
    &2[r_3(x,t)-r_1(x,t)]\frac{E[m(x,t)]}{K[m(x,t)]}
  \end{split}
\end{equation}
where $E(m)$ is the complete elliptic integral of the second kind.

The next step is to write down the first three conservation equations
for the KdV equation \cite{Debnath2005}
\begin{gather}
  u_t + \Big(\tfrac{1}{2} u^2 + \eps^2u_{xx} \Big)_x = 0 \nonumber \\
  \Big(\tfrac{1}{2}u^2 \Big)_t + \Big(\tfrac{1}{3} u^3 + \eps^2 u u_{xx} -
    \tfrac{1}{2} \eps^2 u_x^2 \Big)_x = 0 
  \label{eq:69} \\
  \Big(\tfrac{1}{3} u^3 - \eps^2 u_x^2 \Big)_t\, + \nonumber \\
  \Big(\tfrac{1}{4} u^4 -2\eps^4 u_x u_{xxx} + \eps^4 u_{xx}^2 +
    \eps^2 u^2 u_{xx} - 2\eps^2 u u_{x}^2 \Big)_x = 0 . \nonumber
\end{gather}
We require three equations because there are three parameters
$\{r_i\}$ that are allowed to slowly vary in time and space.  Now we
insert the periodic elliptic function solution $\phi$ into equations
(\ref{eq:69}) and average the equations over the period $L$ to find
\begin{gather*}
  \left(\overline{\phi} \right)_t +
  \Big(\tfrac{1}{2}\overline{\phi^2}\Big)_x = 0 \\ 
  \Big(\tfrac{1}{2}\overline{\phi^2}\Big)_t +
  \Big(\tfrac{1}{3}\overline{\phi^3} - \tfrac{3}{2}
  \overline{\phi_{\theta}^2} \Big)_x = 0 \\
  \Big(\tfrac{1}{3}\overline{\phi^3} - \overline{\phi_{\theta}^2} \Big)_t +
  \Big(\tfrac{1}{4}\overline{\phi^4} - 4 \overline{\phi \phi_{\theta}^2} +
  3\overline{\phi_{\theta}^2} \Big)_x = 0 .
\end{gather*}
Note that $\phi_x = \phi_{\theta}/\eps$.

\begin{figure}
  \centering
  \includegraphics[scale=.5]{
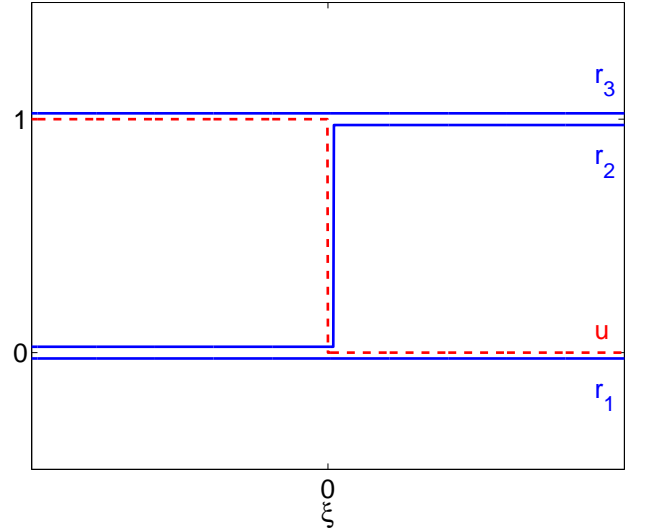}
  \caption{Initial data regularization for the KdV dispersive Riemann
    problem.  The dashed line represents the initial data
    \eqref{eq:14} for $u$.  The solid lines represent the initial data
    for the Riemann invariants $r_1$, $r_2$, and $r_3$ that regularize
    the initial data for $u$.  This initial data for the Riemann
    invariants satisfies the three properties of characterization,
    non-decreasing, and ordering \eqref{eq:47} so a rarefaction wave
    solution exists for all time (see
    Fig. \ref{fig:kdv_ri_evolution}).}
  \label{fig:kdv_initial_data}
\end{figure}

Assuming that the parameters $\{r_i\}$ depend on the slow variables
$x$ and $t$, the above equations can be transformed to Riemann
invariant form \cite{Whitham65,GuPi}
\begin{subequations}
  \label{eq:128}
  \begin{equation}
    \label{eq:31}
    \pd{r_i}{t} + v_i(r_1,r_2,r_3) \pd{r_i}{x} = 0 , \quad i=1,2,3.
  \end{equation}
  The variables $\{r_i\}$ are the Riemann invariants for the hyperbolic
  system (\ref{eq:31}) with the velocities
  \begin{equation}
    \label{eq:127}
    \begin{split}
      v_1 &= \frac{1}{3}(r_1+r_2+r_3) -
      \frac{2}{3}(r_2-r_1)\frac{K(m)}{K(m) - E(m)} \\
      v_2 &= \frac{1}{3}(r_1+r_2+r_3) -
      \frac{2}{3}(r_2-r_1)\frac{(1-m)K(m)}{E(m) -
        (1-m)K(m)} \\
      v_3 &= \frac{1}{3}(r_1+r_2+r_3) -
      \frac{2}{3}(r_3-r_1)\frac{(1-m)K(m)}{E(m)} . 
    \end{split}
  \end{equation}
\end{subequations}
We wish to solve equations \eqref{eq:128} subject to the initial data
\eqref{eq:14}.  This is accomplished by the method of initial data
regularization, presented in \cite{Kodama99}.  Although the background
to this method involves a detailed understanding of inverse spectral
theory and Riemann surface theory, the method itself is
straightforward.  Any solution to equation \eqref{eq:55} with
decreasing initial data will eventually break.  The dissipative
regularization handles this by introducing jump conditions across the
shock relating its speed to its values before and after the
discontinuity.  The dispersive regularization employs the higher order
hyperbolic system \eqref{eq:128} with initial data that
\emph{characterizes} the initial data for $u$, is
\emph{non-decreasing}, and satisfies a \emph{separability} condition.
The initial data
\begin{equation}
  \label{eq:46}
  r_1(x,0) \equiv 0, \quad r_2(x,0) = \left \{
    \begin{array}{cc}
      0 & x\le 0 \\
      1 & x>0 
    \end{array} \right. , \quad r_3(x,0) \equiv 1,
\end{equation}
shown in Fig.\! \ref{fig:kdv_initial_data} regularizes the IVP
\eqref{eq:11} and \eqref{eq:14} because of the following properties
\begin{equation}
  \label{eq:47}
  \begin{array}{cc}
    \overline{\phi}(x,0) = u(x,0;\eps) & \text{(characterization)}, \\[2mm]
    \pd{r_i}{x}(x,0) \ge 0 & \text{(non-decreasing)}, \\[2mm]
    \underset{x \in \real}{\max} \,
    r_i(x,0) < \underset{x \in \real}{\min} \, 
    r_{i+1}(x,0) & \text{(separability)}.
  \end{array}
\end{equation}
Characterization amounts to verifying that the initial data for the
full problem \eqref{eq:14} is equivalent to the initial data for the
averaged problem $\overline{\phi}$; the same assumption is made in the
boundary matching method \cite{GuPi}.  The non-decreasing and
separability of the $r_i$ ensure that a global, continuous
(non-breaking) solution to the Whitham equations \eqref{eq:128} exists
for all time \cite{Levermore88,Lax73}.

\begin{figure}
  \centering
  \includegraphics[scale=.6]{
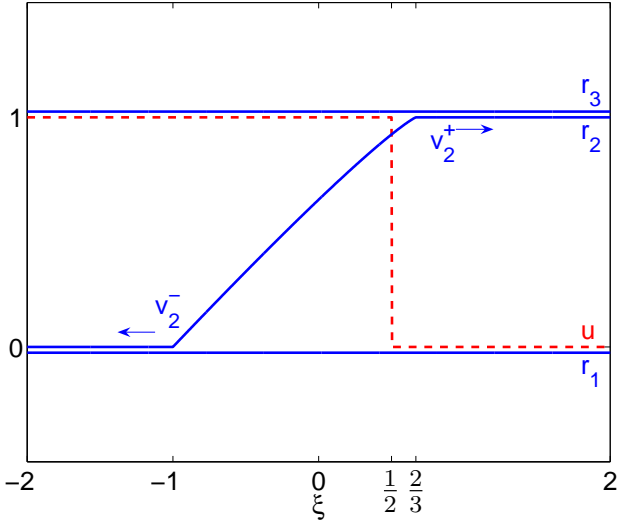}
  \caption{Dissipative (dashed) and dispersive (solid) regularizations
    of the conservation law \eqref{eq:55}.  The dissipative case
    corresponds to a traveling discontinuity with speed $\tfrac{1}{2}$
    satisfying the jump condition \eqref{eq:63}.  The dispersive case
    is a rarefaction wave solution to the Whitham equations
    \eqref{eq:128} with two associated speeds $v_2^+ = 2/3$ and $v_2^-
    = -1$.  This rarefaction wave modulates the periodic solution
    \eqref{eq:28} giving a DSW (see Fig. \ref{fig:gurevich} and
    eq. \eqref{eq:107}).}
  \label{fig:kdv_ri_evolution}
\end{figure}
The system (\ref{eq:128}) with initial data \eqref{eq:46} has an exact
rarefaction solution in the form of a self-similar simple wave with
$r_1 \equiv 0$, $r_3 \equiv 1$, and $r_2(x,t) = r_2(\xi)$, $\xi =
x/t$.  The remaining nontrivial equation in (\ref{eq:31}) takes the
form
\begin{equation*}
  (v_2 - \xi)r_2' = 0,
\end{equation*}
which is satisfied when the implicit relation $v_2 = \xi$ or
\begin{equation}
  \label{eq:57}
  \tfrac{1}{3}[1+r_2(\xi)] - \tfrac{2}{3} r_2(\xi)
  \frac{[1-r_2(\xi)]K[r_2(\xi)]} {E[r_2(\xi)] - [1 -
    r_2(\xi)]K[r_2(\xi)]} = \xi ,
\end{equation}
is satisfied.  The above is one equation for one unknown, $r_2(\xi)$,
which is solved by a standard root finding method for each $\xi$ (see
Fig.\!  \ref{fig:kdv_ri_evolution}).

The rarefaction wave has two associated speeds $v_2^-$ and $v_2^+$
which are determined from the Whitham equations \eqref{eq:128}.  Ahead
of the moving fronts, the $r_i$ are constant.  Since
\begin{equation*}
  \frac{d r_2}{d t} = 0, \quad \text{when} ~ \frac{d x}{d t} = v_2,
\end{equation*}
from equations \eqref{eq:127}, the speeds are given by the limits
\begin{align}
  \label{eq:12}
  v_2^+ &= \lim_{r_2 \to 1^-}v_2(0,r_2,1) = \frac{2}{3}, \\
  \label{eq:58}
  \quad v_2^- &= \lim_{r_2 \to 0^+} v_2(0,r_2,1) = -1.
\end{align}

\begin{figure}
  \centering
  \includegraphics[scale=.4]{
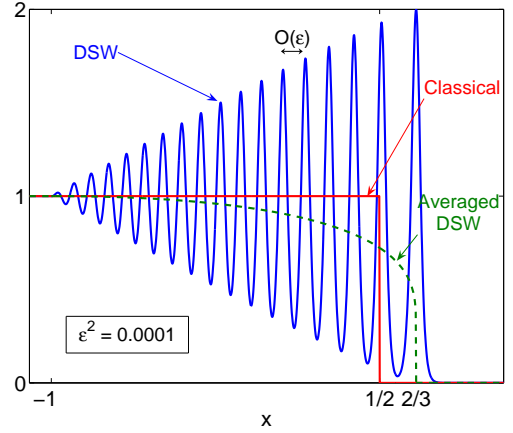}
  \caption{Comparison of a dispersive shock (eq. \eqref{eq:107} with
    $\eps^2 = 0.0001$), its average $\overline{\phi}$
    (eq. (\ref{eq:26}), dashed), and a classical, dissipative shock
    (eq. \eqref{eq:53}, Burgers' solution) plotted at time $t = 1$.
    The DSW front moves faster than its classical counterpart.  The
    average $\overline{\phi}$ looks very similar to the classical
    shock, a steep front connected to a constant in the rear, except
    that the speed of the front is different and the function is
    continuous.}
  \label{fig:gurevich}
\end{figure}
The dispersive Riemann problem, equation \eqref{eq:11} with initial
data \eqref{eq:14} or, equivalently, equations \eqref{eq:28} and
\eqref{eq:128} with initial data \eqref{eq:46}, has the asymptotic ($t
\gg 1$ and $\eps^2 \ll 1$) DSW solution
\begin{equation}
  \label{eq:107}
  \begin{split}
    u(x,t;\eps) &\approx r_2(x/t) - 1 + 2\text{dn}^2\!\left( \frac{x -
        V(x/t) t}{\eps\sqrt{6}};m = r_2(x/t) \right) \\
    V(x/t) &= \tfrac{1}{3}(1 + r_2(x/t)).
  \end{split}
\end{equation}
The function $r_2(x/t)$ is the rarefaction wave solution satisfying
equation \eqref{eq:57}.  This DSW solution, its average
(eq. (\ref{eq:26})), and the Burgers type classical shock solution
(\ref{eq:53}) are shown in Fig.\!  \ref{fig:gurevich}.  The DSW
averaging process produces a shock front that resembles the classical
shock front but, however, has a different speed and the DSW front is
continuous.  The front speed of the DSW, $\tfrac{2}{3}$, is the phase
speed of the classical soliton solution to KdV which fixes an
amplitude of 2
\begin{equation*}
  u(x,t) = 2 \sech^2 \! \Big(\tfrac{1}{\sqrt{6\eps^2}}(x - \tfrac{2}{3}t)\Big).
\end{equation*}
One can think of a DSW as a slowly modulated train of solitons
decaying, in a self-similar fashion, to a constant.  The DSW is based
on the rarefaction solution \eqref{eq:57} so it has two associated
speeds, the trailing edge $v_2^-$ \eqref{eq:58} and the leading edge
$v_2^+$ \eqref{eq:12}.  Even though the DSW is non-zero as $x \to
-\infty$, the oscillations remain in a finite region of space
describing the expanding behavior of a steep gradient in this
dispersive system.

\begin{figure}
  \centering
  \includegraphics[scale=.47]{
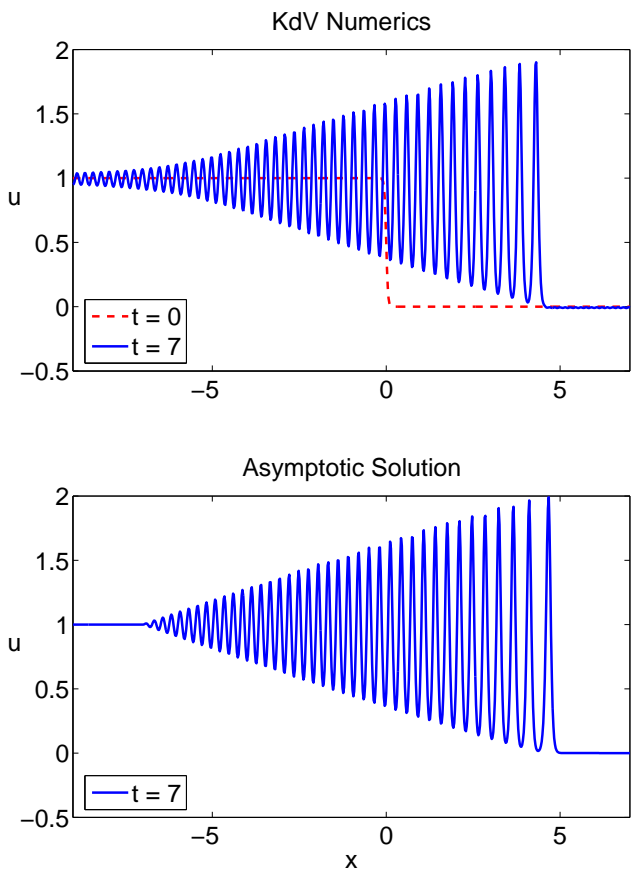}
  \caption{Numerical solution of the KdV equation with an initial
    step (top) and the asymptotic DSW solution \eqref{eq:107}
    (bottom) for $\eps^2 = 0.001$.}
  \label{fig:asymp_compare_kdv}
\end{figure}

In Fig. \ref{fig:asymp_compare_kdv}, we show the numerical solution to
the KdV equation with the step initial data \eqref{eq:14} for $\eps^2
= 0.001$.  The wavelength of oscillation, leading edge amplitude, and
speed of the asymptotic solution agree well with the numerical result.
The position of the leading edges differ slightly because the
asymptotic solution is valid for $t \gg 1$ and it takes the numerical
solution some time to reach this stage.  Note that the speed of the
leading edge in the numerical solution, averaged from $t = 5$ to $t =
7$, is 0.660 which is approximately $\tfrac{2}{3}$, the analytical
result \eqref{eq:12}.

\begin{figure}
  \centering
  \includegraphics[scale=.45]{
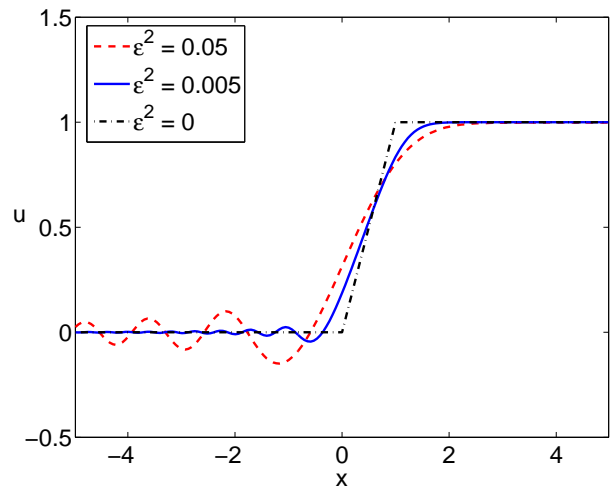}
  \caption{Numerical solution of the KdV equation \eqref{eq:11} with
    the step initial data \eqref{eq:4} for different values of $\eps$
    and the zero dissipation/dispersion limit $\eps = 0$, the
    rarefaction wave \eqref{eq:5}.  The plot corresponds to $t = 1$.}
  \label{fig:kdv_rarefaction}
\end{figure}

If the initial data for the KdV equation \eqref{eq:11} is
non-decreasing then no breaking occurs and a global solution exists
for the zero dispersion limit ($\eps^2 \to 0$).  For the case
\begin{equation}
  \label{eq:4}
  u(x,0;\eps) = \left \{
    \begin{array}{cc}
      0 & x<0 \\
      1 & x>0
    \end{array} \right. ,
\end{equation}
the solution is a rarefaction wave, a weak solution satisfying the
conservation law \eqref{eq:55}.  This is the Burgers rarefaction wave
\eqref{eq:5} (see Fig. \ref{fig:kdv_rarefaction}).

Whitham averaging provides an effective way to define the DSW shock
speeds and derive the asymptotic oscillatory structure of a DSW along
with its leading amplitude.  We note that in any experiment $\eps$
will be finite hence oscillations will exist.  The averaged solution
is useful when comparing with gas dynamics since we can evaluate the
jump in density across a DSW region and determine the velocity of a
DSW shock front.

The long time asymptotic behavior of the KdV equation has been
analyzed in \cite{AbSe77}.  In general, an arbitrary initial condition
will evolve into a dispersive tail \cite{AbSe77}, a set of solitons
\cite{GaGrKrMi67,GaGrKrMi74}, and a "collisionless shock" region
\cite{AbSe77}.  In a sense, the asymptotic solution in Fig.
\ref{fig:gurevich} contains all of these regions.  The very front of
the oscillations is the collisionless shock region over which a
constant connects to a train of $\sech^2$ solitons eventually leading
to small, linear oscillations at the tail.

Thus we have described how to study a dispersive shock wave associated
with KdV in the context of Whitham theory.  A DSW can arise in the
dispersive regularization of a conservation law just as a classical
shock can arise in the dissipative regularization of a conservation
law.  The key difference is that a \emph{weak limit} where one
averages over the oscillations is required in the dispersive case.
This method gives useful results such as the asymptotic modulated
oscillatory profile, the wavelength of oscillation, the leading
amplitude, and the speeds of a dispersive shock.  On a large scale,
once the limiting process has been accomplished, the DSW and classical
shock look similar, i.e. constants connected by sharp gradients.
However, the shock speeds are different.

 On a small mathematical note, Lax and Levermore
\cite{LaLe83All} used the inverse scattering transform to take the
limit $\eps \rightarrow 0$ in the KdV equation \eqref{eq:11} for a
broad class of initial data.  They showed that the limit,
$\overline{u}$, is a weak limit in the sense that
\begin{equation*}
  \lim_{\eps \to 0} \int_{-\infty}^{\infty}u(x,t;\eps) f(x) \, \ud x =
  \int_{-\infty}^{\infty}\overline{u}(x,t) f(x) \, \ud x
\end{equation*}
for all smooth, compactly defined functions $f(x)$.  This type of
limiting procedure is required because the solution develops an
infinite number of oscillations.  Lax and Levermore also showed that,
in a region of breaking, the weak limit $\overline{u}$ can be
calculated explicitly by using the Whitham averaging method thus
giving the method a strong mathematical footing. 

Throughout the rest of this paper, we will use this
dissipative/dispersive analogy with the Burgers' and KdV equations to
motivate our discussion of the more complicated problem involving the
dissipative and dispersive regularizations arising in the context of
BEC and gas dynamics.

\subsection{Dissipative Regularization of the Euler Equations}
\label{sec:visc-regul}
As mentioned in the introduction, the compressible equations of gas
dynamics without dissipation are the same as the local conservation
equations for a BEC \eqref{eq:19} with $\eps=0$.  Let us consider the
Riemann problem for the dissipative regularization of the Euler
equations in one dimension with step initial data
\begin{equation}
  \label{eq:20}
  \begin{split}
    \rho_t + (\rho u)_x &= 0 \\
    (\rho u)_t + (\rho u^2 + \tfrac{1}{2}\rho^2)_x &= 0 \\[2mm]
    \rho(x,0) = \left\{
      \begin{array}{lc}
        \rho_0 & x<0 \\
        1 & x>0
      \end{array}
    \right. ~, \quad
    &u(x,0) = \left\{
      \begin{array}{lc}
        u_0 & x<0 \\
        0 & x>0
      \end{array}
    \right.  . 
  \end{split}
\end{equation}
This is a general step initial value problem with two parameters
$\rho_0$ and $u_0$.  Note if we make the transformation
\begin{equation}
  \label{eq:38}
  \begin{split}
  \tilde{\rho} = \rho_r \rho~, \quad \tilde{u} = \sqrt{\rho_r} u +
  u_r~, \\
  t = \rho_r \tilde{t}~, \quad x = \sqrt{\rho_r} (
  \tilde{x} - u_r \tilde{t} ), \quad \rho_r \ne 0,
  \end{split}
\end{equation}
then we convert the initial conditions in (\ref{eq:20}) to the general
step initial conditions
\begin{align*}
  \tilde{\rho}(\tilde{x},0) &= \left\{
    \begin{array}{lr}
      \rho_l=\rho_r \rho_0 & \tilde{x}<0 \\
      \rho_r & \tilde{x}>0
      \end{array}
    \right. \\
    \tilde{u}(\tilde{x},0) &= \left\{
      \begin{array}{lr}
        u_l=\sqrt{\rho_r}u_0 + u_r & \tilde{x}<0 \\
        u_r & \tilde{x}>0
      \end{array}
    \right. .
\end{align*}

The unique weak solution to (\ref{eq:20}) consists of, in general,
three constant states connected to one another via two wave solutions:
shocks or centered rarefaction waves (this is a special case of
\eqref{eq:22} for $n = 2$).  We will first study the canonical,
right-going shock case which consists of two constant states connected
by a traveling discontinuity.

The jump conditions \eqref{eq:63} for the IVP \eqref{eq:20} are,
assuming a traveling wave shock,
\begin{equation}
  \label{eq:39}
  \begin{split}
    V(\rho_0 - 1) &= \rho_0 u_0 \\
    V(\rho_0 u_0) &= \rho_0 u_0^2 + \tfrac{1}{2}\rho_0^2 -
    \tfrac{1}{2} . 
  \end{split}
\end{equation}
A physically realizable shock solution must satisfy an entropy
condition, the statement that across any shock wave, there must be a
corresponding increase in entropy.  For the Riemann problem
(\ref{eq:20}), the entropy condition for a right going shock is simply
$\rho_0 > 1$ \cite{Le02}.  
The entropy condition and the jump conditions (\ref{eq:39})
determine a classical shock wave uniquely.  Solving for $u_0$ and $V$
in (\ref{eq:39}) gives
\begin{equation}
  \label{eq:43}
  u_0 = \pm (\rho_0 - 1)\sqrt{\tfrac{1}{2} (1+ \frac{1}{\rho_0}) },
\end{equation}
with the corresponding shock speed
\begin{equation*}
  V = \frac{\rho_0 u_0}{\rho_0 - 1}. 
\end{equation*}
For an entropy satisfying right-going shock, we take the plus sign in
(\ref{eq:43}).  With this specific choice for $u_0$, the Riemann
problem (\ref{eq:20}) has the unique weak solution, parameterized by
$\rho_0$,
\begin{equation*}
  \rho(x/t) = \left\{
    \begin{array}{lr}
      \rho_0 & x/t<V \\
      1 & x/t>V
    \end{array}
  \right. ,\quad
    u(x/t) = \left\{
      \begin{array}{lr}
        u_0 & x/t<V \\
        0 & x/t>V
      \end{array}
    \right . ,
\end{equation*}
a shock moving with speed
\begin{equation*}
  V = \rho_0 \sqrt{\frac{\rho_0 - 1/\rho_0}{2(\rho_0 - 1)}}. 
\end{equation*}
Note that for a right-going shock to exist, there must be a non-zero
density $\rho$ on the right.  Otherwise, the solution is purely a
rarefaction wave \cite{Le02}.

When $\rho_0 < 1$, a pure rarefaction solution exists for the specific
choice
\begin{equation*}
  u_0 = 2(\sqrt{\rho_0}-1).
\end{equation*}
This choice is a consistency condition for the existence of a
continuous rarefaction wave connecting the left constant state
$(\rho_0,\rho_0 u_0)$ and the right state $(1,0)$ \cite{Le02}.  The
rarefaction solution is given by (also see
Fig. \ref{fig:nls_rarefaction})
\begin{equation}
  \label{eq:106}
  \begin{split}
    \rho(x/t) &= \left \{
      \begin{array}{cc}
        \rho_0 & x/t < (3\sqrt{\rho_0}-2) \\[1mm]
        \tfrac{1}{9}(2+x/t)^2 & (3\sqrt{\rho_0}-2) < x/t < 1 \\[1mm]
        1 & 1 < x/t
      \end{array} \right. \\[2mm]
    u(x/t) &= \left \{
      \begin{array}{cc}
        2\sqrt{\rho_0}-2 & x/t < (3\sqrt{\rho_0}-2) \\[1mm]
        \tfrac{1}{3}(-2+2x/t)^2 & (3\sqrt{\rho_0}-2) < x/t < 1 \\[1mm]
        0 & 1 < x/t
      \end{array} \right. .
  \end{split}
\end{equation}

By manipulation, the Euler equations \eqref{eq:20} can be written in
the Riemann invariant form 
\begin{equation}
  \label{eq:8}
  \begin{split}
    \pd{r_+}{t} + \tfrac{1}{4}(r_+ + 3r_-)\pd{r_+}{x} &= 0 \\
    \pd{r_-}{t} + \tfrac{1}{4}(3r_+ + r_-)\pd{r_-}{x} &= 0 \\[2mm]
    r_+ = u + 2\sqrt{\rho}, \quad r_- = u - &2\sqrt{\rho} \\[2mm]
  \end{split}
\end{equation}
These equations yield a general solution of the form \eqref{eq:22}.
In the next section, we will compare the solution of equations
\eqref{eq:8} to the zero dispersion limit of the GP equation.

The dissipative regularization of the Euler equations--the jump
conditions \eqref{eq:39}--gives a criterion for determining the speed
of a classical shock wave given the initial jump in density.  
In the next section, we will discuss DSWs in BEC and show that
rarefaction waves play a crucial role in their understanding.

\subsection{Dispersive Regularization of the Euler Equations, BEC}
\label{sec:disp-regul}
Consider equations \eqref{eq:19}, the local conservation equations for
a BEC, with $\eps^2 \to 0$.  This is the zero dispersion limit of the
GP equation, which we consider as the dispersive regularization of the
Euler equations considered in the previous section.  Assuming free
expansion of the condensate or zero potential, $V =0$, the GP equation
is equivalent to the 3D, defocusing nonlinear Schr\"{o}dinger equation
(NLS).  Later we show that the 1D NLS equation gives good qualitative
agreement with the three dimensional problem in the so-called "blast
wave" regime. In this regard, the Whitham averaging method is a useful
device to analyze DSWs in a BEC. It has been used to analyze the 1D
defocusing NLS equation in the case of an initial jump in density and
velocity using boundary matching \cite{GuKr87,El95,KaGaKr04}.  We
apply Whitham averaging to the 1D NLS equation using initial data
regularization as presented in \cite{Kodama99,BiondiniKodama05} to
derive the shock structure of the canonical 1D BEC DSW along with its
associated speeds.  This technique is equivalent to the boundary
matching method for a single dispersive shock wave.  It also allows
for generalizations to more complicated, multi-phase type interactions
which we will study in the future.

For small dispersion, oscillations begin to develop in breaking
regions.  As in the KdV case, a strong limit does not exist; hence we
are lead to consider a weak limit where one averages over the
oscillations.  Consider the dispersive Riemann problem for BEC in one
dimension without an external potential $V$, which models a freely
expanding condensate
\begin{equation}
  \label{eq:21}
  \begin{split}
    \rho_t + (\rho u)_x &= 0 \\
    (\rho u)_t + (\rho u^2 + \frac{1}{2} \rho^2)_x &= \frac{\eps^2}{4}
    \Big( \rho \big(\log \rho \big)_{xx} \Big)_x \\
    \rho(x,0;\eps) = \left\{
      \begin{array}{lc}
        \rho_0 & x<0 \\
        1 & x>0
      \end{array}
    \right. ~&, \quad
    u(x,0; \eps) = \left\{
      \begin{array}{lc}
        u_0 & x<0 \\
        0 & x>0
      \end{array}
    \right.  ~. 
  \end{split}
\end{equation}
Recall that $\rho = |\Psi|^2$ represents the condensate density and $u
= \eps (\arg\, \Psi)_x$ is the condensate flow velocity.  These
initial conditions are general step-like data (see \eqref{eq:38}).

To apply Whitham theory, we require a periodic solution to equation
\eqref{eq:21}, the local conservation equations for the 1D NLS
equation
\begin{equation}
  \label{eq:108}
  i\eps \Psi_t + \tfrac{1}{2}\eps^2 \Psi_{xx} - |\Psi|^2 \Psi = 0.
\end{equation}
Assume a traveling wave solution of the form
\begin{equation*}
  \Psi(x,t) = A(\theta) e^{i \phi(\theta)/\eps}, \quad \theta = (x-Vt)/\eps.
\end{equation*}
Inserting this ansatz into equation \eqref{eq:108}, and equating real
and imaginary parts gives
\begin{equation}
  \label{eq:110}
  \begin{split}
    -V A' +\phi' A' + \tfrac{1}{2} \phi'' A &= 0 \\
    V\phi' A + \tfrac{1}{2} A'' - \tfrac{1}{2} \phi'^2 A - A^3 &=
    0.
  \end{split}
\end{equation}
Integrating the first equation and solving for $\phi'$, we find
\begin{equation*}
  \phi' = V - \frac{2c_1}{A^2},
\end{equation*}
where $c_1$ is a constant of integration.  Inserting this result into
the second equation in \eqref{eq:110} and simplifying gives
\begin{equation*}
  A'' + V^2 A - \frac{4 c_1^2}{A^3} - 2 A^3 = 0.
\end{equation*}
Integrating this equation gives
\begin{equation}
  \label{eq:113}
  A'^2 + V^2 A^2 + \frac{4c_1^2}{A^2} - A^4 + c_2 = 0,
\end{equation}
where $c_2$ is a second constant of integration.  To obtain the
elliptic function solution, let $\rho = A^2$; then equation
\eqref{eq:113} becomes
\begin{equation}
  \label{eq:114}
  \begin{split}
    \rho'^2 &= 4(\rho^3 - V^2 \rho^2 - 2 c_2 \rho - 4 c_1^2) \\
    &= 4(\rho - \lambda_1)(\rho - \lambda_2)(\rho - \lambda_3), \\[2mm]
    & 0 < \lambda_1 < \lambda_2 < \lambda_3.
  \end{split}
\end{equation}
The periodic solution to the above equation is \cite{ByrdFriedman54}
\begin{equation}
  \label{eq:115}
  \begin{split}
    \rho(x,t) &= \lambda_3 - (\lambda_3 - \lambda_1)
    \text{dn}^2(\sqrt{\lambda_3-\lambda_1}\frac{(x-Vt)}{\eps};m) \\
    u(x,t) &= \phi'(\theta) = V - \frac{2 c_1}{\rho(x,t)}, ~ V =
    \sqrt{\lambda_1 + \lambda_2 + \lambda_3} \\
    m &= \frac{\lambda_2-\lambda_1}{\lambda_3-\lambda_1}, \quad c_1^2
    = \tfrac{1}{4} \lambda_1 \lambda_2 \lambda_3.
  \end{split}
\end{equation}
Similar to the KdV equation, this solution has three independent
constants of integration $\lambda_i$, $i = $ 1, 2, 3, the roots of the
cubic polynomial in \eqref{eq:114}.  However, by the invariance of the
NLS equation with respect to the ``Galilean boost''
\begin{equation*}
  \Psi(x,t) \to e^{-i\tilde{V}(x-\tfrac{1}{2}\tilde{V}t)}
  \Psi(x-\tilde{V}t,t),
\end{equation*}
the phase speed $\tilde{V}$ is another arbitrary constant.  Then the
periodic solution to equation \eqref{eq:21} is
\begin{equation}
  \label{eq:40}
  \begin{split}
    \rho(x,t;\eps) &\equiv \psi(\theta) = \lambda_3 - (\lambda_3 -
    \lambda_1) \text{dn}^2(\sqrt{\lambda_3-\lambda_1}\theta;m)
    \\
    u(x,t;\eps) &\equiv \nu(\theta) = V - \sigma\frac{\sqrt{\lambda_1 \lambda_2
        \lambda_3}}{\psi(\theta)} \\
    m &= \frac{\lambda_2-\lambda_1}{\lambda_3-\lambda_1} , \quad
    \sigma = \pm 1, \quad \theta = \frac{x-Vt}{\eps},
  \end{split}
\end{equation}
with four arbitrary parameters $V$ and $\lambda_i$, $i=1,2,3$.  The
sign of the constant of integration $c_1$ in \eqref{eq:115} is not
determined.  Either sign $\sigma = \pm 1$ gives a valid periodic
solution to the NLS equation \eqref{eq:108}.  This will be important
later in our analysis of DSWs with points of zero density.

Using Whitham's averaging method, we will derive equations describing
slow modulations of the four parameters in the periodic solution
\eqref{eq:40}.  Anticipating the form of the Whitham equations, we
will express the four aforementioned parameters in terms of four
parameters $r_i$ \cite{GuKr87}
\begin{equation}
  \label{eq:41}
  \begin{split} 
    V &= \tfrac{1}{4}(r_1+r_2+r_3+r_4), \\
    \lambda_3 &= \tfrac{1}{16}(-r_1-r_2+r_3+r_4)^2 ,\\
    \lambda_2 &= \tfrac{1}{16}(-r_1+r_2-r_3+r_4)^2, \\
    \lambda_1 &= \tfrac{1}{16}(r_1-r_2-r_3+r_4)^2, \\[2mm]
    &r_1 < r_2 < r_3 < r_4.
  \end{split}
\end{equation}
Similar to the KdV system, the parameters $r_i$ correspond to Riemann
invariants for the NLS Whitham equations, which we derive now.

The period of the dn function in \eqref{eq:40} is $[\sqrt{\lambda_3 -
  \lambda_1} \theta] = 2K(m)$ where $[\cdot]$ denotes the period of
the argument.  Then the wavelength of oscillation for the periodic
solution \eqref{eq:40} is
\begin{equation}
  \label{eq:78}
  L = \frac{2K(m)}{\sqrt{\lambda_3 - \lambda_1}}.
\end{equation}
The average of the periodic solution \eqref{eq:40} over the wavelength
$L$ is \cite{ByrdFriedman54}
\begin{equation}
  \label{eq:79}
  \begin{split}
    \overline{\psi}(x,t) &= \frac{1}{L} \int_0^L \psi(\theta) \,
    \ud \theta \\
    &= \lambda_3  - (\lambda_3 - \lambda_1) \frac{E(m)}{K(m)} \\
    \overline{\psi \nu}(x,t) &= \frac{1}{L} \int_0^L \psi(\theta)
    \nu(\theta) \, \ud \theta \\
    &= V \overline{\psi} - \sigma \sqrt{\lambda_1 \lambda_2
      \lambda_3}, \\
    \overline{\nu}(x,t) &=  \frac{1}{L} \int_0^L \nu(\theta) \,
    \ud \theta \\
    &= V - \sigma\sqrt{\frac{\lambda_1 \lambda_2}{\lambda_3}} -
    \sigma\sqrt{\lambda_3 - \lambda_1}\{E(\chi,1-m) + \\[1mm]
    &\qquad\qquad ~ F(\chi,1-m)[E(m)/K(m) - 1]\} \\[2mm]
    \chi &= \sin^{-1}\!\!\left( \sqrt{\frac{\lambda_3-\lambda_1}{\lambda_3}}
      \right) 
  \end{split}
\end{equation}
The functions $F(\chi,1-m)$ and $E(\chi,1-m)$ are incomplete elliptic
integrals of the first and second kind respectively
\cite{ByrdFriedman54}.

 Four conservation laws for the NLS equation are
\cite{JiLeMc94,AblowitzSegurSolitons} 
\begin{equation}
  \label{eq:76}
  \begin{split}
    \rho_t + (\rho u)_x = 0& \\
    (\rho u)_t + \Big(\rho u^2 + \frac{1}{2} \rho^2 - \frac{\eps^2}{4}
    \rho \big(\log \rho \big)_{xx} \Big)_x = 0& \\
    \Big(\rho u^2 + \rho^2 + \eps^2 \frac{\rho_x^2}{4\rho} \Big)_t +
    \Big(\rho u^3
    + 2 \rho^2 u + \eps^2 \frac{\rho_x^2 u}{4 \rho} \Big)_x = 0& \\
    \Big( \eps^3\rho_{xxx} - \tfrac{3}{2} \eps^3\frac{\rho_x
      \rho_{xx}}{\rho} + \tfrac{3}{4}\eps^3\frac{\rho_x^3}{\rho^2} -
    5\eps\rho \rho_x - 3 \eps\rho_x u^2 + & \\
    3\eps\rho u u_x \Big)_t +
    \Big( \tfrac{1}{2}\eps^4\rho_{xxxx} - \tfrac{5}{4} \eps^4
    \frac{\rho_{xxx}\rho_x}{\rho} - \tfrac{3}{4} \eps^2
    \frac{\rho_{xx}}{\rho} +& \\
    \tfrac{21}{8} \eps^4 \frac{\rho_{xx}
      \rho_x^2}{\rho^2} - \tfrac{7}{2}\eps^2 \rho \rho_{xx} -
    \tfrac{9}{2} \eps^2u^2 \rho_{xx}  - \tfrac{9}{8}
    \eps^4\frac{\rho_x^4}{\rho^3} 
    - \tfrac{1}{4}\eps^2 \rho_x^2 + & \\
    3\eps^2 \frac{u^2 \rho_x^2}{\rho} -
    - 6\eps^2uu_x\rho_x +
    2\rho^3 +7\rho^2u^2 - & \\
    5\eps^2\rho u u_{xx} -
    3\eps^2\rho u_x^2 + 2\rho u^4 +
    2\rho^2 u^2\Big)_x = 0&
  \end{split}
\end{equation}
To obtain the Whitham equations for the $\{r_i\}$, we insert the
periodic solution \eqref{eq:40} into the conservation laws
\eqref{eq:76} and average over the fast variable $\theta$ to find
\begin{equation}
  \label{eq:75}
  \begin{split}
    (\overline{\psi})_t + \Big(\overline{\psi \nu} \Big)_x = 0& \\
    \Big( \overline{\psi \nu} \Big)_t + \Big( \overline{\psi \nu^2} +
    \tfrac{1}{2} \overline{\psi^2} - \tfrac{1}{4} \overline{\psi (\log
      \psi )_{\theta \theta}} \Big)_x = 0& \\
    \Big(\overline{\psi \nu^2} + \overline{\psi^2} +
    \overline{\frac{\psi_{\theta}^2}{4\psi}} \Big)_t +
    \Big(\overline{\psi \nu^3} + 2 \overline{\psi^2 \nu} +
    \overline{\frac{\psi_{\theta}^2 \nu}{4
        \psi}} \Big)_x = 0& \\
    \Big(     \tfrac{3}{4}\overline{\frac{\psi_\theta^3}{\psi^2} } 
    - \tfrac{3}{2} \overline{\frac{\psi_\theta
        \psi_{\theta\theta}}{\psi}} 
    - \tfrac{9}{2} \overline{\psi_\theta \nu^2} \Big)_t +
    \Big(- \tfrac{5}{4} 
    \overline{\frac{\psi_{\theta\theta\theta}\psi_\theta}{\psi}} - \tfrac{3}{4} 
    \overline{\frac{\psi_{\theta\theta}}{\psi}} +& \\
    \tfrac{21}{8}  \overline{\frac{\psi_{\theta\theta}
      \psi_\theta^2}{\psi^2}} - \tfrac{7}{2} \overline{\psi
    \psi_{\theta\theta}} - 
    \tfrac{9}{2} \overline{\nu^2 \psi_{\theta\theta}}  - \tfrac{9}{8}
    \overline{\frac{\psi_\theta^4}{\psi^3} }
    - \tfrac{1}{4} \overline{\psi_\theta^2} + 
    3 \overline{\frac{\nu^2 \psi_\theta^2}{\psi} }
    + & \\
    2\overline{\psi^3} +7\overline{\psi^2\nu^2} +
    \overline{\psi \nu \nu_{\theta\theta}} +
    3\overline{\psi \nu_\theta^2} + 2\overline{\psi \nu^4} +
    2\overline{\psi^2 \nu^2} \Big)_x = 0&
  \end{split}
\end{equation}
 Assuming that the four parameters $r_i$ vary on the slow length
and time scales $x$ and $t$, the Whitham equations are obtained
\cite{Pavlov87,GuKr87}
\begin{subequations}
  \label{eq:131}
  \begin{equation}
    \label{eq:42}
    \pd{r_i}{t} + v_i(r_1,r_2,r_3,r_4) \pd{r_i}{x} = 0 , \quad i=1,2,3,4. 
  \end{equation}
  The $v_i$ are expressions involving complete first and second elliptic
  integrals
  \begin{equation}
    \label{eq:77}
    \begin{split}
      v_1 &= V - \tfrac{1}{2}(r_2 - r_1)\left[1 - \frac{(r_4-r_2)E(m)}
        {(r_4-r_1)K(m)} \right]^{-1}\\
      v_2 &=  V + \tfrac{1}{2}(r_2 - r_1)\left[1 - \frac{(r_3-r_1)E(m)}
        {(r_3-r_2)K(m)} \right]^{-1}\\
      v_3 &= V - \tfrac{1}{2}(r_4 - r_3)\left[1 - \frac{(r_4-r_2)E(m)}
        {(r_3-r_2)K(m)} \right]^{-1}\\
      v_4 &= V + \tfrac{1}{2}(r_4 - r_3)\left[1 - \frac{(r_3-r_1)E(m)}
        {(r_4-r_1)K(m)} \right]^{-1} \\
      m &= \frac{(r_4-r_3)(r_2-r_1)}{(r_4-r_2)(r_3-r_1)} .
    \end{split}
  \end{equation}
\end{subequations}
Thus (the complicated looking) equations \eqref{eq:75} reduce to the
simple system of first order PDE \eqref{eq:131}.  This reduction takes
advantage of certain integrable properties of the NLS equation which
we will not discuss here.  For further details, the reader is referred
to \cite{ForestLee86}.

As in the KdV case, it turns out that the right going dispersive shock
wave is characterized by a self-similar, simple rarefaction wave in
the associated Whitham equations \eqref{eq:131}.  There are two free
parameters in the IVP \eqref{eq:21}, the initial velocity $u_0$ and
the initial density $\rho_0$.  As we are seeking a simple wave
solution, we require one of the Riemann invariants from the Euler
system \eqref{eq:8}, $r_-$ in this case, to be constant initially.
For the initial data in \eqref{eq:21}, this corresponds to the
specific choice for the initial velocity $u_0 = 2(\sqrt{\rho_0}-1)$
and the initial data
\begin{equation}
  \label{eq:71}
  r_+(x,0) = \left \{
    \begin{array}{cc}
      4\sqrt{\rho_0} - 2 & x<0 \\
      2 & x>0
    \end{array} \right., ~ r_-(x,0) \equiv -2.
\end{equation}
Equations \eqref{eq:8} accurately describe a regular solution, hence a
dispersive or dissipative regularization of the Euler equations
whenever both $r_+$ and $r_-$ are non-decreasing \cite{Kodama99}.  A
weak limit is not required because a rarefaction type solution exists
for all time.  In other words, the dissipative and dispersive
regularizations are equivalent when no shocks develop.

On the other hand, when the initial data for the Riemann invariants
are decreasing, as in the case of \eqref{eq:71} with $\rho_0 > 1$, a
shock wave will develop.

\begin{figure}
  \centering
  \includegraphics[scale=.55]{
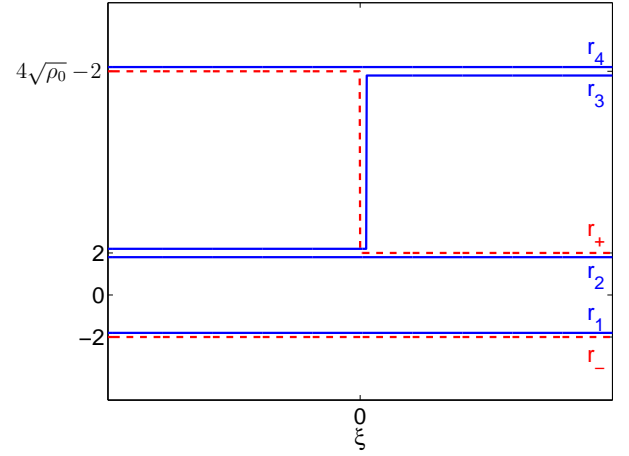}
  \caption{Regularized initial data for the dispersive Riemann problem
    \eqref{eq:21}.  The variables $r_+$ and $r_-$ are the Riemann
    invariants for the Euler equations satisfying \eqref{eq:8}.  The
    $\{r_i\}$ are the Riemann invariants for the Whitham equations
    \eqref{eq:131} describing the modulation of a periodic wave
    \eqref{eq:40}.  The $\{r_i\}$ satisfy the properties of being
    non-decreasing and separable \eqref{eq:86}, so a continuous
    rarefaction solution exists for all time (see
    Fig. \ref{fig:dsw_ri_evolution}).}
  \label{fig:dsw_initial_data}
\end{figure}

We now use the technique of initial data regularization
\cite{Kodama99} where the $r_i$ are chosen initially (see
Fig. \ref{fig:dsw_initial_data}) so that
\begin{equation}
  \label{eq:86}
  \begin{array}{cc}
    \overline{\psi}(x,0) = \rho(x,0;\eps), ~ \overline{\nu}(x,0) =
    u(x,0;\eps) & \text{(characterization)}, \\[2mm]
    \pd{r_i}{x}(x,0) \ge 0 & \text{(non-decreasing)}, \\[2mm]
    \underset{x \in \real}{\max}\, 
    r_i(x,0) < \underset{x \in \real}{\min} \, 
    r_{i+1}(x,0) & \text{(separability)}.
  \end{array}
\end{equation}
Recall that the characterization property implies that the initial
data for the full problem and the averaged problem are the same.  For
this to be so, we take
\begin{equation}
  \label{eq:129}
  \begin{split}
    r_1(x,0) &\equiv -2, \quad r_2(x,0) \equiv 2, \quad r_4(x,0)
    \equiv 4\sqrt{\rho_0} - 2, \\
    r_3(x,0) &= \left\{
      \begin{array}{cc}
        2 & x<0 \\
        4\sqrt{\rho_0} - 2 & x>0
      \end{array}
      \right. .
  \end{split}
\end{equation}
We also require a spatial dependence of the sign $\sigma$ in
\eqref{eq:79} when $\rho_0 \ge 4$
\begin{equation}
  \label{eq:120}
  \sigma(x,0) = \sgn(x) = \left \{
    \begin{array}{cc}
      -1 & x<0 \\
      1 & x>0
    \end{array} \right. .
\end{equation}
When $1 \le \rho_0 < 4$, $\sigma \equiv 1$.  The non-decreasing and
separability properties guarantee that a continuous solution for the
hyperbolic system \eqref{eq:131} exists for all time
\cite{Kodama99,Lax73}.  Thus we have regularized the ``shock'' initial
data.

\begin{figure}
  \centering
  \includegraphics[scale=.55]{
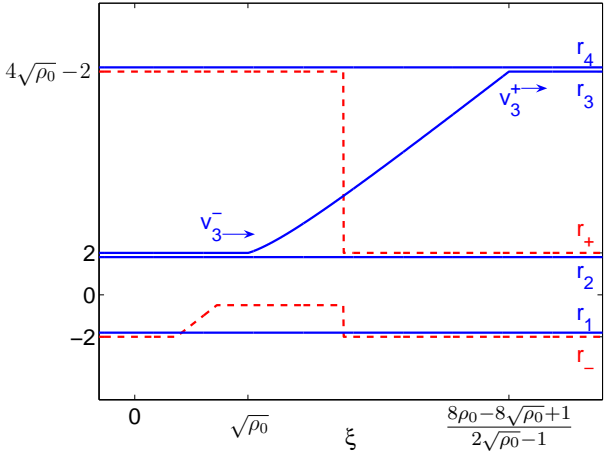}
  \caption{Solution to classical (dashed) and dispersive (solid)
    Riemann problems for the Euler equations with initial data shown
    in Fig. \ref{fig:dsw_initial_data}.  The classical solution
    consists of three constant states connected by a rarefaction wave
    and a shock wave. The dispersive regularization involves a pure
    rarefaction solution of the Whitham equations \eqref{eq:131} where
    only $r_3$ varies according to \eqref{eq:48}.  This solution
    modulates the periodic solution \eqref{eq:40} giving a DSW (see
    Fig. \ref{fig:nls_dsw}).}
  \label{fig:dsw_ri_evolution}
\end{figure}
Using the initial data regularization shown in
Fig. \ref{fig:dsw_initial_data}, we solve equations \eqref{eq:131} for
a self-similar ($\xi = x/t$), simple rarefaction wave.  Assuming $r_1
= -2$, $r_2 = 2$, $r_3(x,t) = r_3(\xi)$, and $r_4 = 4\sqrt{\rho_0} -
2$, we find that all the Whitham equations are satisfied if
\begin{equation*}
  (v_3 - \xi)r_3' = 0.
\end{equation*}
This equation is satisfied for non-constant $r_3$ when $v_3 = \xi$, or
explicitly
\begin{equation}
  \label{eq:48}
  \begin{split}
    \xi &= \tfrac{1}{4} r_3(\xi) + \sqrt{\rho_0} - \tfrac{1}{2} -
    [4\sqrt{\rho_0} - 2 - r_3(\xi)] \times \\[1.5mm]
    & \qquad \qquad \times \left[ 2 - \frac{(8 \sqrt{\rho_0} -
        8)E(m(\xi))}{(r_3(\xi) -2)K(m(\xi))} \right]^{-1} \\[3mm]
    m(\xi) &= \frac{4\sqrt{\rho_0} - 2 - r_3(\xi)}{(\sqrt{\rho_0} -
      1)(r_3(\xi) + 2)}.
  \end{split}
\end{equation}
The above is one nonlinear equation for the unknown $r_3(\xi)$. We use
a standard root finding method to solve this system.  The rarefaction
solution for $r_3$ is shown in Fig. \ref{fig:dsw_ri_evolution} along
with the dissipative regularization of the Euler equations
\eqref{eq:8} for the same initial data.  Recall that the general
solution for the dissipative Riemann problem has the form
\eqref{eq:22} for $n = 2$.  For the initial data in \eqref{eq:71}, we
find that this general solution consists of three constant states, the
first two connected by a rarefaction wave, the latter two connected by
a dissipative shock wave (see the dashed curve in
Fig. \ref{fig:dsw_ri_evolution}).

\begin{figure}
  \centering
  \includegraphics[scale=.45]{
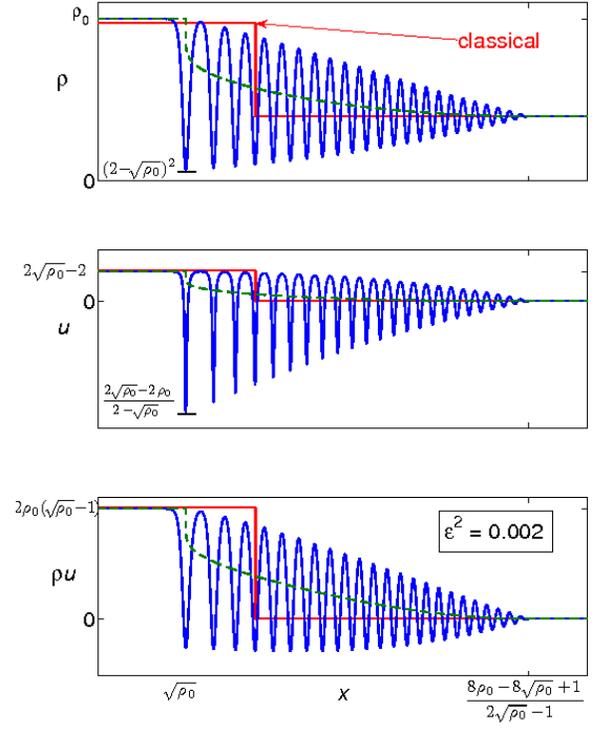}
  \caption{DSW (GP, oscillatory) and classical shock (Euler,
    discontinuity).  The dashed curve is the averaged DSW.  The upper
    plot is the density $\rho$, the middle plot is the velocity $u$,
    and the lower plot is the momentum $\rho u$ for the parameters
    $\rho_0 = 2.5$, $t = 1$, and $\eps^2 = 0.002$.  A DSW in BEC is an
    expanding region with the constant trailing and leading edge
    speeds $v_3^-=\sqrt{\rho_0}$ and $v_3^+=\frac{8\rho_0 -
      8\sqrt{\rho_0} + 1}{2\sqrt{\rho_0} - 1}$ respectively.  When $1
    < \rho_0 < 4$, the DSW looks similar to the one pictured here.
    For $\rho_0 \ge 4$, the situation is different see
    Fig. \ref{fig:vacuum_point}.  The maximums and minimums of the
    density and velocity are marked for comparison with the NLS gray
    soliton solution \eqref{eq:49}.}
  \label{fig:nls_dsw}
\end{figure}

Inserting the self-similar solution of the Whitham equations
\eqref{eq:48} into the original periodic solution (\ref{eq:40}) for
the case $1 \le \rho_0 < 4$ (recall $\sigma \equiv 1$), gives a slowly
modulated periodic wave (see Fig. \ref{fig:nls_dsw}), a DSW in BEC.

\begin{figure}
  \centering
  \includegraphics[scale=.45]{
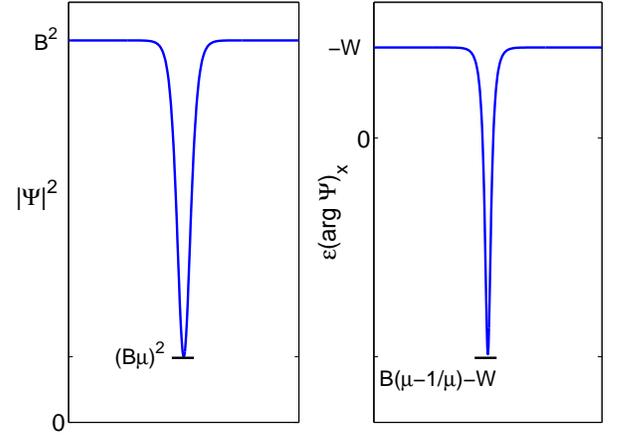}
  \caption{The density and velocity of the gray soliton solution
    \eqref{eq:49} with labeled maximum and minimum values for
    comparison with the trailing dip in the DSW
    Fig. \ref{fig:nls_dsw}.}
  \label{fig:gray}
\end{figure}

The speeds of the trailing and leading edges of the DSW are found in
the same manner as they were in the KdV case, using equations
\eqref{eq:77}
\begin{align}
  \label{eq:29}
  v_3^- &= \lim_{r_3 \to 2^+}
  v_3(-2,2,r_3,4\sqrt{\rho_0}-2) = \sqrt{\rho_0}, \\
  v_3^+ &= \lim_{r_3 \to 4\sqrt{\rho_0}-2^-}
  v_3(-2,2,r_3,4\sqrt{\rho_0}-2) \nonumber \\
  \nonumber
  & = \frac{8\rho_0 - 8 
    \sqrt{\rho_0} + 1}{2\sqrt{\rho_0} - 1}.
\end{align}
Because the trailing edge speed is the speed of sound (linear
disturbances of the Euler equations \eqref{eq:20}) in gas dynamics
(see e.g. \cite{Le02}), it might appear that the trailing edge of a
DSW is not affected by the nonlinearities in the problem.  However a
closer examination reveals that the trailing edge speed is exactly the
phase speed of a gray soliton with the trailing dip shape given in
Fig. \ref{fig:nls_dsw}.  A general gray soliton solution to the 1D NLS
equation \eqref{eq:108} is
\begin{equation}
  \label{eq:49}
  \begin{split}
    &\Psi(x,t) = B e^{-i[(B^2 + \frac{1}{2}W^2)t + Wx]/\eps} [\beta
    \tanh(\theta) + i\mu ], \\
    &\rho = |\Psi(x,t)|^2 = B^2[\beta^2\tanh^2 (\theta) + \mu^2 ], \\
    &u = \eps [\arg \Psi(x,t)]_x = -W - \frac{\mu B \beta^2
      \sech^2(\theta)}{B^2 \tanh^2(\theta) + \mu^2}    \\
    & \theta = B \beta[x - (B\mu - W)t - x_0]/\eps.
\end{split}
\end{equation}
where all parameters are real and $\mu^2 + \beta^2 = 1$ 
(see Fig. \ref{fig:gray}). The speed of the gray soliton is
\begin{equation}
  \label{eq:50}
  v_{g} = B\mu - W. 
\end{equation}
It's minimum density and velocity occur when $x = (B\mu - W)t + x_0$
giving the maximum and minimum values
\begin{equation}
  \label{eq:105}
  \begin{split}
    \rho_{max} &= \max |\Psi|^2 = B^2, \quad \rho_{min} = \min
    |\Psi|^2 = (B\mu)^2,\\
    u_{max} &= \max \eps (\arg \Psi)_x = -W,\\
    u_{min} &= \min \eps (\arg \Psi)_x = B(\mu - \frac{1}{\mu}) - W.
  \end{split}
\end{equation}
To find the parameters $B$, $\mu$, and $W$ we relate the maximums and
minimums of the general gray soliton in \eqref{eq:105} to the maximums
and minimums of the trailing dip in the DSW of Fig. \ref{fig:nls_dsw}.
Equating maxes and mins for the densities, we find
\begin{equation}
  \label{eq:104}
  B = \sqrt{\rho_0}, \quad \mu  = \frac{2 -
    \sqrt{\rho_0}}{\sqrt{\rho_0}}.
\end{equation}
Setting the maximum velocity equal, $-W = 2\sqrt{\rho_0} - 2$, we find
\begin{equation}
  \label{eq:51}
  W = 2 - 2\sqrt{\rho_0}. 
\end{equation}
So the phase speed \eqref{eq:50} of the gray soliton with the
parameters \eqref{eq:104} and \eqref{eq:51} is
\begin{equation*}
  v_g = \sqrt{\rho_0},
\end{equation*}
equivalent to \eqref{eq:29}, the DSW trailing edge speed.  The
trailing edge of the DSW moves with the phase speed of a gray soliton
which, in this case, is the sound speed.  Therefore, similar to the
KdV case and, as argued in \cite{KaGaKr04}, the trailing edge of the
DSW can be thought of as a modulated train of gray solitons.

\begin{figure}
  \centering
  \includegraphics[scale=.45]{
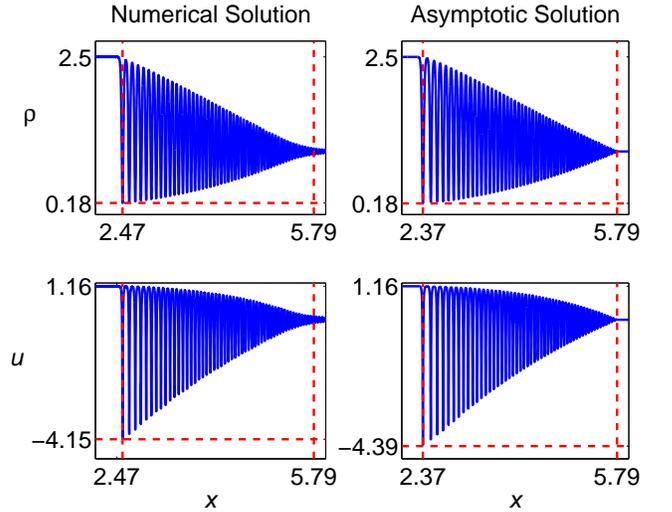}
  \caption{Numerical solution of the dispersive Riemann problem (left)
    and its asymptotic solution (right) at $t = 1.5$ for $\rho_0 =
    2.5$ and $\eps = 0.03$.  The numerically determined trailing edge
    speed, calculated from $t = 1.2$ to $t = 1.5$, is 1.590,
    approximately equal to $\sqrt{\rho_0} = 1.581$, the theoretical
    result \eqref{eq:29}.  The trailing edge density is 0.175,
    approximately the theoretical value $(2-\sqrt{\rho_0})^2 =
    0.184$.}
  \label{fig:nls_asymp_compare}
\end{figure}

In Fig. \ref{fig:nls_asymp_compare}, the asymptotic result depicted in
Fig. \ref{fig:nls_dsw} is compared with direct numerical simulation of
the NLS equation with step initial data showing excellent agreement.

\begin{figure}
  \centering
  \includegraphics[scale=0.45]{
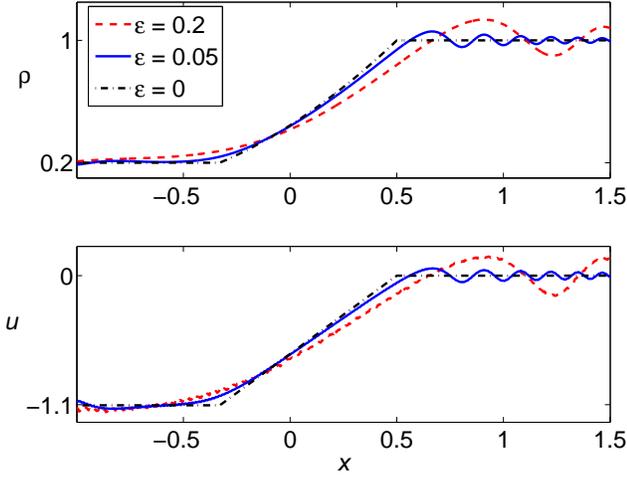}
  \caption{Numerical solution of the dispersive Riemann problem for
    rarefaction initial data.  As $\eps \to 0$, the dispersive
    regularization converges strongly to the dissipative
    regularization, the rarefaction wave eq. \eqref{eq:106}.  The
    parameters are $\rho_0 = 0.2$, $u_0 = 2\sqrt{\rho_0}-2$, and $t =
    0.5$.}
  \label{fig:nls_rarefaction}
\end{figure}

The shock profile in Fig. \ref{fig:nls_dsw} is valid when $1 < \rho_0
< 4$.  When $\rho_0 < 1$, the initial data for the Riemann invariants
$r_+$ and $r_-$ \eqref{eq:71} is non-decreasing therefore no initial
data regularization is required and a rarefaction wave solution
exists, see eq.  \eqref{eq:106} and Fig. \ref{fig:nls_rarefaction}.
The rarefaction solution for the dispersive and dissipative
regularizations is the same.

For the case $\rho_0 \ge 4$, a point of the solution with zero
density, a vacuum point, is generated \cite{El95}.  We label the
location of the vacuum point with the similarity variable $\xi_v$.
The minimum density of $\rho(x,t)$ in \eqref{eq:40} is $\rho_{min} =
\lambda_1$ (since $\min( -\text{dn}(x)^2) = -\text{dn}(0)^2 = -1$).
Solving $\rho_{min} = \lambda_1 = 0$ at a vacuum point with the
constants in \eqref{eq:129} ($r_1 = -2$, $r_2 = 2$, $r_4 =
4\sqrt{\rho_0}-2$) and using equation \eqref{eq:41}
($r_1-r_2-r_3+r_4=0$) gives $r_3(\xi_v) = 4\sqrt{\rho_0} - 6$.  Using
the relation \eqref{eq:48}, the location is
\begin{equation}
  \label{eq:33}
  \begin{split}
    \xi_v &= 2\sqrt{\rho_0} - 2 - 2\left[ 1 - \frac{(\sqrt{\rho_0} -
        1)E(m_v)} {(\sqrt{\rho_0} - 2)K(m_v)}  \right]^{-1} \\
    m_v &= (\sqrt{\rho_0} - 1)^{-2}.
  \end{split}
\end{equation}

When $\rho_0 \ge 4$, the characterization of the initial data in
\eqref{eq:86} requires that the sign $\sigma$ in \eqref{eq:40} change
at the origin \eqref{eq:120}.  Since the modulated periodic solution
\eqref{eq:40} and its average \eqref{eq:79} depend on $\sigma$, we
must determine how $\sigma$ depends on $x$ and $t$.  It is natural to
assume that the globally conserved momentum $\rho u =
i\frac{\eps}{2}(\Psi \Psi^*_x - \Psi^*\Psi_x)$ is smooth.  The
modulated periodic wave for the momentum is (recall \eqref{eq:40})
\begin{equation*}
  \rho u = V\rho - \sigma \sqrt{\lambda_1 \lambda_2 \lambda_3}.
\end{equation*}
Since $V$ and $\rho$ are smooth, we consider
\begin{equation*}
  \begin{split}
    \sigma \sqrt{\lambda_1 \lambda_2 \lambda_3} = &\frac{1}{64}
    \sigma |4\sqrt{\rho_0} - 6 - r_3(\xi)| \times\\
    &(4\sqrt{\rho_0} - 2 -
    r_3(\xi))(4\sqrt{\rho_0} - 6 + r_3(\xi)).
  \end{split}
\end{equation*}
The above expression is derived from the rarefaction solution
\eqref{eq:48} and the relations \eqref{eq:41}.  Then, for $\sigma
\sqrt{\lambda_1 \lambda_2 \lambda_3}$ to be smooth, we require
\begin{equation*}
  \sigma(\xi) = \sgn(4\sqrt{\rho_0} - 6 - r_3(\xi)).
\end{equation*}
Since 
\begin{equation*}
  r_3(\xi_v) = 4\sqrt{\rho_0} - 6,
\end{equation*}
the sign change occurs exactly at the vacuum point $\xi = \xi_v$
\eqref{eq:33}.  Thus, $\sigma$ has the self-similar dependence
\begin{equation}
  \label{eq:3}
  \sigma(x,t) = \sigma(x/t) = \sgn(x/t-\xi_v) = \left \{
    \begin{array}{cc}
      -1 & x/t < \xi_v \\
      1 & x/t > \xi_v
    \end{array} \right. .
\end{equation}

\begin{figure}
  \centering
  \includegraphics[scale=.45]{
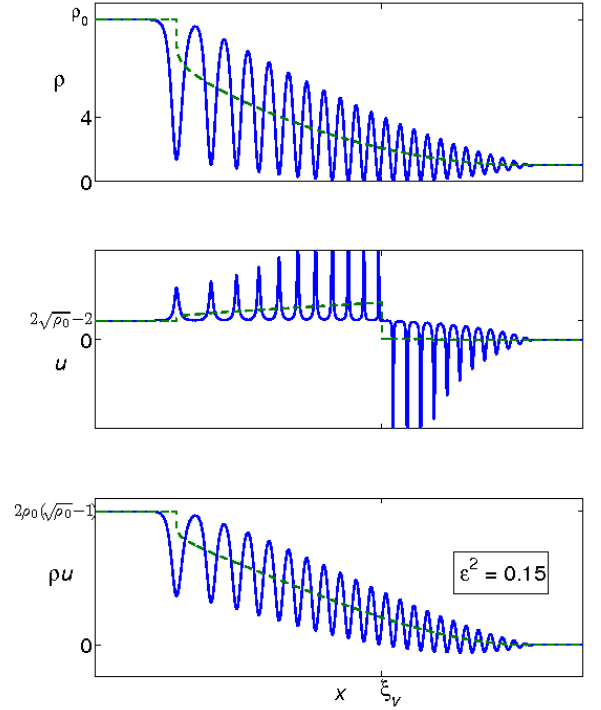}
  \caption{DSW shock profile (solid) and its average (dashed) plotted
    at $t = 1$ when $\rho_0 \ge 4$ (here $\rho_0 = 10$) giving rise to
    a vacuum point at $x/t = \xi_v \approx 7.4$ (see
    eq. \eqref{eq:33}) of zero density and infinite velocity.  The
    average velocity $\overline{\nu}$ exhibits a jump at the vacuum
    point but the average momentum $\overline{\phi \nu}$ does not.
    This indicates that the variables $\rho$ and $\rho u$ are the most
    natural variables for the problem.}
  \label{fig:vacuum_point}
\end{figure}

An example DSW profile and its average with a vacuum point at $x/t =
\xi_v$, are shown in Fig. \ref{fig:vacuum_point}.  At the vacuum point
there is a jump in the average velocity $\overline{\nu}$.  However,
there is no jump in the average momentum $\overline{\psi \nu}$.

\begin{figure}
  \centering
  \includegraphics[scale=.45]{
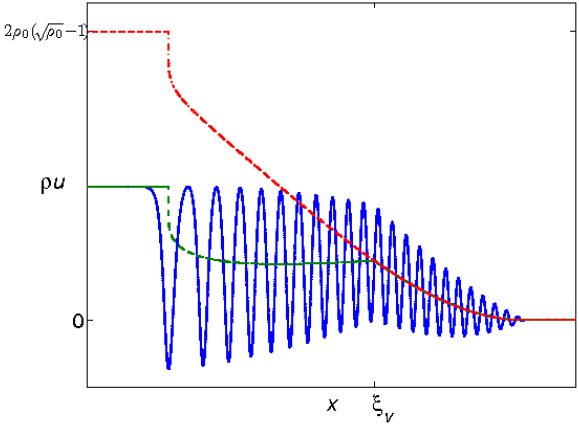}
  \caption{Average momentum $\overline{\psi \nu}$ \eqref{eq:79}
    (dashed) and the modulated DSW momentum $\psi \nu$ \eqref{eq:40}
    (solid) with the rarefaction solution \eqref{eq:48} and the
    \emph{incorrect} choice of constant sign $\sigma \equiv +1$ and
    the \emph{correct} choice $\sigma = \sgn(\xi - \xi_v)$ (dash
    dotted) for $\rho_0 = 10$.  The boundary condition \eqref{eq:126}
    is not satisfied and there is a kink at the vacuum point for the
    incorrect $\sigma$.}
  \label{fig:wrong_sign}
\end{figure}

For comparison, Fig. \ref{fig:wrong_sign} is a plot of the average
momentum $\overline{\psi\nu}$ (eq. \eqref{eq:79}) and the modulated
DSW momentum $\psi \nu$ (eq. \eqref{eq:40}) with the \emph{incorrect}
choice $\sigma(x,t) \equiv +1$.  The trailing edge condition
\begin{equation}
  \label{eq:126}
  \lim_{x\to-\infty}
  \overline{\psi\nu} = 2\rho_0(\sqrt{\rho_0}-1)
\end{equation}
is \emph{not} satisfied.  Also, the average momentum is not smooth at
the vacuum point $x/t = \xi_v$.

From the plot of the velocity $u$ in Fig. \ref{fig:vacuum_point}, we
will argue in the next section that vacuum points appear in the blast
wave experiments considered.

In summary, the asymptotic solution for a DSW in a one-dimensional BEC
is a slowly modulated periodic wave expressed as
\begin{equation}
  \label{eq:2}
  \begin{split}
    \rho(x,t;\eps) &\approx \lambda_3(x/t) - [\lambda_3(x/t) -
    \lambda_1(x/t)] \times \\
    & \qquad \quad \times
    \text{dn}^2[\sqrt{\lambda_3(x/t)-\lambda_1(x/t)}\theta;m(x/t)],
    \\[2mm]
    u(x,t;\eps) &\approx V(x/t) -
    \sigma(x/t)\frac{\sqrt{\lambda_1(x/t) \lambda_2(x/t)
        \lambda_3(x/t)}}{\rho(x,t;\eps)}, \\
    m(x/t) &=
    \frac{\lambda_2(x/t)-\lambda_1(x/t)}{\lambda_3(x/t)-\lambda_1(x/t)}
    , \\
    V(x/t) &= \tfrac{1}{4}[4\sqrt{\rho_0}-2 + r_3(x/t) ],
    \quad \theta = \frac{x-V(x/t)t}{\eps}, \\
    \lambda_1(x/t) &= \tfrac{1}{16}[4\sqrt{\rho_0} - 6 -
    r_3(x/t)]^2, \\
    \lambda_2(x/t) &= \tfrac{1}{16}[4\sqrt{\rho_0} + 2 -
    r_3(x/t)]^2, \\
    \lambda_3(x/t) &= \tfrac{1}{16}[4\sqrt{\rho_0} - 2 + r_3(x/t)]^2 ,
  \end{split}
\end{equation}
where $r_3$ satisfies the implicit equation \eqref{eq:48}. For $1 <
\rho_0 < 4$, $\sigma(x/t) \equiv 1$.  When $\rho_0 \ge 4$, there is a
vacuum point and $\sigma$ is given in equation \eqref{eq:3}.

Here we find that a key difference between dispersive and dissipative
shock waves is the method of regularization.  In the dispersive case,
we use initial data regularization.  In particular we argue that, with
the correct choice of initial data, the hyperbolic Whitham equations
have a smooth solution for all time.  Whereas in the dissipative case,
jump/entropy conditions are employed.  Using a dispersive
regularization, we have determined the behavior of a fundamental DSW
in Bose-Einstein condensates.  The averaged behavior of the DSW is
similar to the classical shock case but the speeds and oscillatory
behavior are different.

\subsection{Theoretical Explanation of Experiments}
\label{sec:theor-expl-exper}
\begin{figure}
  \centering
  \begin{tabular}{cc}
    $\quad|\Psi|^2\qquad\qquad$ & $\qquad\qquad\qquad\quad(\arg \Psi)_r$
  \end{tabular}
  \includegraphics[scale=.45]{
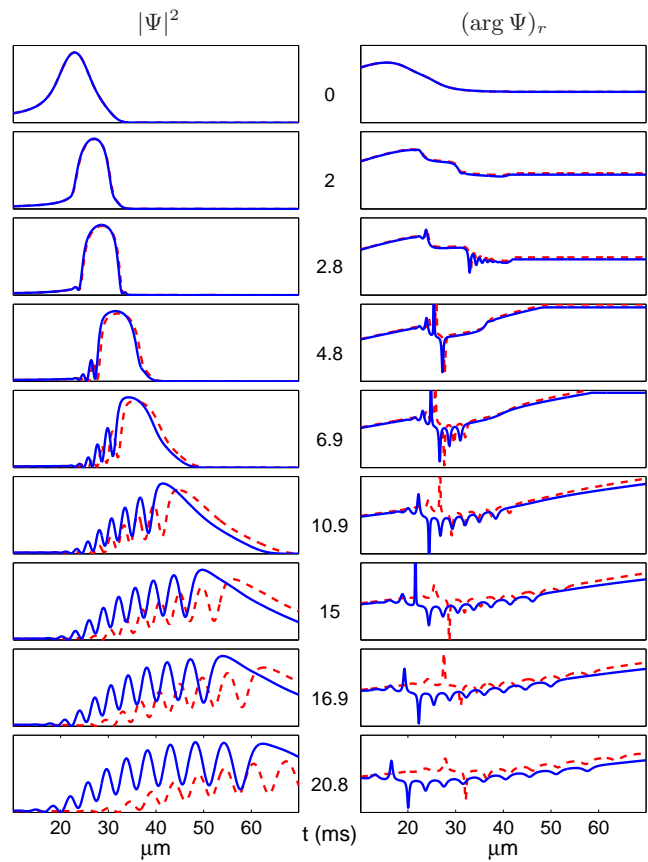}
  \caption{Development and propagation of DSW in the 3D in trap
    experiment with (dashed) and without (solid) the expansion
    potential.  The expansion potential does not alter the structure
    of the DSW, just the speeds.  The 3D and 2D densities are in
    excellent agreement with a maximum relative error of less than
    1\%.  The difference cannot be seen in this graphic.}
  \label{fig:3d_shock_intrap}
\end{figure}

In order to investigate the development of dispersive shock waves in
2D and 1D, we assume that the condensate is ``prepared'' by the 3D
evolution, using the results of the two experiments with the
potentials $V_{it}$ (eq. \eqref{eq:64}) and $V_{ot}$
(eq. \eqref{eq:66}).  The state of the condensate at a specific time,
$t = \tilde{t}$ (described later), is used as an initial condition for
a new set of equations in one and two dimensions
\begin{eqnarray}
  \text{2D:}\quad \Psi_{\text{2D}}(\rho,t=0) &=&
  \Psi(\rho,z=0,t=\tilde{t}) \nonumber \\
  \label{eq:37}
  \text{1D:}\quad \Psi_{\text{1D}}(x,t=0) &=&
  \Psi(x,z=0,t=\tilde{t}), ~ x \ge 0 \quad \\
  \Psi_{\text{1D}}(-x, t=0) &=& \Psi_{\text{1D}}(x, t=0) .
  \nonumber
\end{eqnarray}
Because the anti-trap term in the potentials $V_{it}$ and $V_{ot}$
mainly serves to speed up condensate expansion, we neglect this term
($\alpha_r = 0$) when comparing the differing dimensional problems.
In Fig. \ref{fig:3d_shock_intrap} the difference between $\alpha_r \ne
0$ and $\alpha_r = 0$ can be seen.  Specifically, we solve the
following three equations numerically in three, two, and one
dimensions respectively {\setlength\arraycolsep{1pt}
\begin{eqnarray}
  i \eps \pd{\Psi_{\text{3D}}}{t} &=& -\frac{\eps^2}{2} \left(
    \pdd{\Psi_{\text{3D}}}{r} + \frac{1}{r}
    \pd{\Psi_{\text{3D}}}{r} + \pdd{\Psi_{\text{3D}}}{z} \right) 
  \nonumber\\
  \nonumber
  & & + \frac{1}{2} \alpha_z z^2
  \Psi_{\text{3D}} + |\Psi_{\text{3D}}|^2 \Psi_{\text{3D}} \\[3mm]
  \label{eq:116}
  i \eps \pd{\Psi_{\text{2D}}}{t} &=& -\frac{\eps^2}{2} \left(
    \pdd{\Psi_{\text{2D}}}{r} + \frac{1}{r}
    \pd{\Psi_{\text{2D}}}{r} \right) + |\Psi_{\text{2D}}|^2
  \Psi_{\text{2D}} \qquad \\[3mm]
  \nonumber
  i \eps \pd{\Psi_{\text{1D}}}{t} &=& -\frac{\eps^2}{2}
  \pdd{\Psi_{\text{1D}}}{x} + |\Psi_{\text{1D}}|^2 \Psi_{\text{1D}},
\end{eqnarray}}
\!with the initial conditions given in \eqref{eq:37} and
$\Psi_{\text{3D}}(r,z,t=0) = \Psi(r,z,t=\tilde{t})$. 

\begin{figure}
  \centering
  \begin{tabular}{cc}
    $\qquad\qquad|\Psi|^2$ &
    $\qquad\qquad\qquad\qquad(\arg 
    \Psi)_r~\text{or}~ (\arg 
    \Psi)_x$
  \end{tabular}
  \includegraphics[scale=.45]{
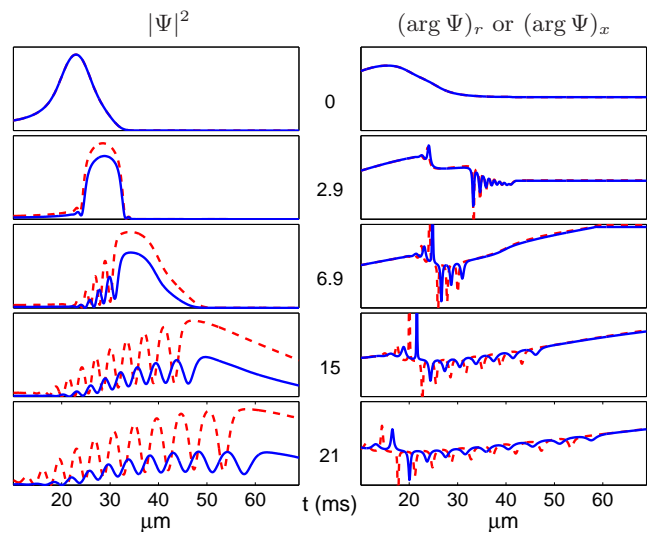}
  \caption{Comparison of 3D $\Psi_{3D}(r,z=0,t)$ (solid) and 1D
    $\Psi_{1D}(x,t)$ (dashed) simulations of DSW in the modified in
    trap experiment showing good qualitative agreement.}
  \label{fig:1d_3d_shock_intrap}
\end{figure}

\begin{figure}
  \centering
  \begin{tabular}{cc}
    $\qquad\qquad|\Psi|^2$ &
    $\qquad\qquad\qquad\qquad(\arg 
    \Psi)_r~\text{or}~ (\arg 
    \Psi)_x$
  \end{tabular}
  \includegraphics[scale=.45]{
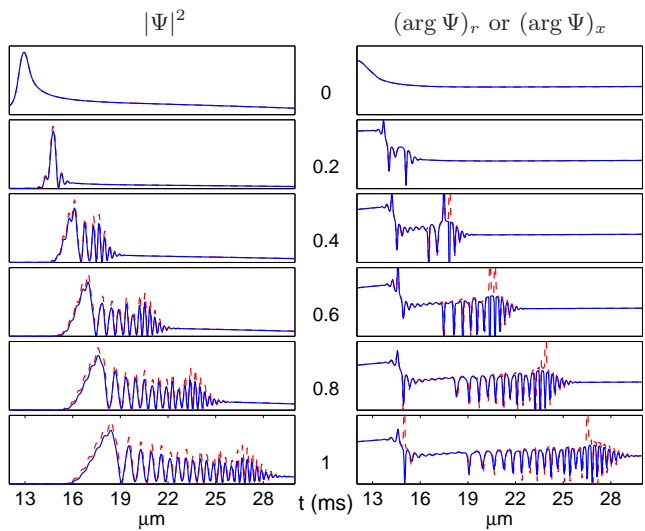}
  \caption{Comparison of 3D (solid) and 1D (dashed) shock development.
    The left plot shows evolution of density; the right plot shows the
    evolution of the velocity.  There is good qualitative agreement
    over this short time scale.}
  \label{fig:1d_2d_shock}
\end{figure}

Comparison of the numerical simulations in different dimensions is
made by considering the density $|\Psi|^2$ and the appropriate
velocity of $\Psi$.  In 3D the radial velocity is defined to be $[\arg
\Psi_{\text{3D}}(r,0,t)]_r$.  In 2D, the velocity is $[\arg
\Psi_{\text{2D}}(r,t)]_r$ whereas in 1D, the velocity is $[\arg
\Psi_{\text{1D}}(x,t)]_x$.  The 3D and 2D results are found to be
barely distinguishable with less than 1\% difference between them in
density, the difference between them cannot be seen in
Fig. \ref{fig:3d_shock_intrap}.  Also, in
Figs. \ref{fig:1d_3d_shock_intrap} and \ref{fig:1d_2d_shock}, it is
shown that the three dimensional and one dimensional results are
qualitatively the same hence the analytical studies in 1D in section
\ref{sec:disp-regul} are reasonable approximations of the 3D case.

First we consider the in trap simulation with the potential $V_{it}$
(eq. \eqref{eq:64}).  We take the state of the condensate at time
$\tilde{t} = \delta t$ (5 ms), just after the laser has been applied.
Now, the evolution is governed solely by the GP equation with an
expansion potential.  Figure \ref{fig:3d_shock_intrap} is a comparison
between the 3D evolution with ($\alpha_r = 0.71$) and without
($\alpha_r = 0$) the expansion potential.  The trailing edge of the
DSW propagates towards the center when there is no expansion potential
whereas in the full simulations with the expansion potential, both the
leading and trailing edges of the DSW propagate radially outward.
Otherwise, the two simulations are very similar.  In order to
investigate dispersive shock behavior, we neglect the expansion
potential, $\alpha_r = 0$ from now on.  We refer to this case as the
modified in trap experiment.

The evolution in Fig. \ref{fig:3d_shock_intrap} shows the development
of a DSW on the inner ring of high density and another DSW on the
outer edge of the ring.  The density is given on the left while the
radial velocity is plotted on the right.  The outer DSW (see $t = 2.8$
ms) loses its strength and vanishes.  The DSW shock structure
investigated in the previous section can clearly be seen in the radial
velocity plot (compare with the asymptotic solution in
Fig. \ref{fig:vacuum_point}).  The appearance of a vacuum point is
clear in the velocity plot as well.

In Fig. \ref{fig:1d_3d_shock_intrap}, the 3D simulation of the
modified in trap experiment is compared with the corresponding 1D
simulation.  Both depict the generation of DSWs and we see that they
are in good qualitative agreement.  The 1D density grows to about
twice the magnitude of the 3D density.  In the velocity plot, the 3D
DSW speeds (trailing and leading) are faster than the 1D speeds.
Therefore, the 1D analysis performed in section \ref{sec:disp-regul}
is applicable to this 3D experiment only in a qualitative sense--it
explains the basic structure of a DSW in 3D.

\begin{figure}
  \centering
  \begin{tabular}{cc}
    $\qquad|\Psi|^2~\text{or}~\rho\quad$ &
    $\qquad\qquad\qquad\quad\eps(\arg 
    \Psi)_r~\text{or}~ u$
  \end{tabular}
  \includegraphics[scale=.45]{
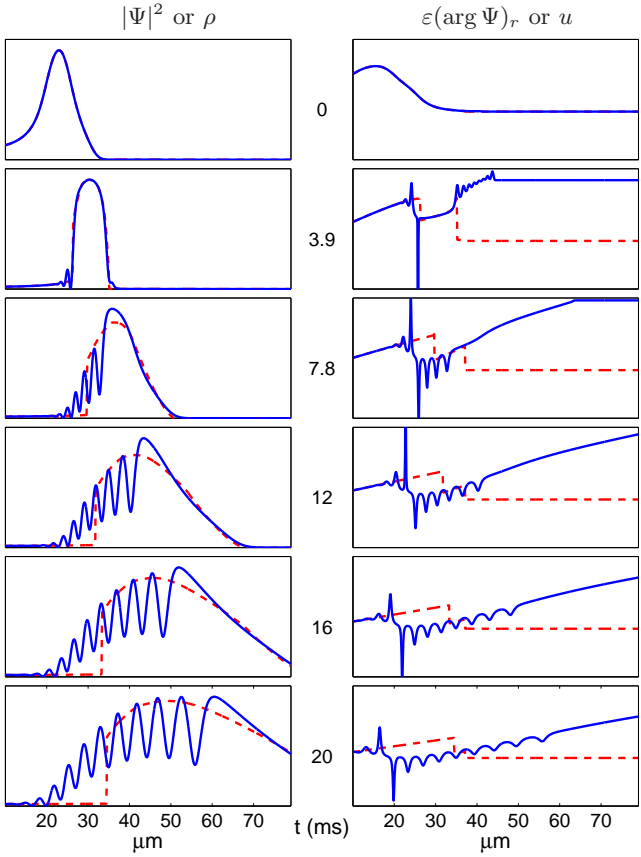}
  \caption{Comparison of 2D simulations for BEC (solid) and gas
    dynamics (dashed).  The dissipative regularization used in the gas
    dynamics simulation was calculated using Clawpack \cite{Le02}.}
  \label{fig:2d_hydro_shock_intrap}
\end{figure}

As outlined in section \ref{sec:class-disp-shock}, dissipative and
dispersive shock waves have quite different properties.  In
Fig. \ref{fig:2d_hydro_shock_intrap}, 2D simulations of the Euler
equations of gas dynamics and a BEC are compared.  The dissipative
regularization of the Euler equations was calculated using the finite
volume package Clawpack \cite{Le02} for the conservation laws
\begin{equation}
  \label{eq:118}
  \begin{split}
    \rho_t + (\rho u)_r + \frac{\rho u}{r} &= 0 \\[2mm]
    (\rho u)_t + ( \rho u^2 + \tfrac{1}{2}\rho^2)_r + \frac{\rho
      u^2}{r} &= 0.
  \end{split}
\end{equation}
The same initial conditions for both the dissipative and dispersive
cases were used.  Both simulations depict the generation of two shock
waves, one on the inner edge of the high density ring and another on
the outer edge.  The outer shock vanishes in both cases as it
propagates into the region of negligible density.  Recall that a shock
wave can only exist when there is a non-zero density on both of its
sides \cite{Le02}. As we have shown, the dissipative and dispersive
shocks propagate with different speeds.  However, the densities in the
two simulations do agree in regions not affected by breaking, the
smooth region to the right of the DSW.  As discussed in the previous
section, the dissipative and dispersive regularizations are the same
when there is no breaking.

For the out of trap experiment, we consider the state of the
condensate at time $\tilde{t}$ corresponding to about 10 ms during the
experiment or 1 ms after the laser has been turned on.  This
particular $\tilde{t}$ was chosen because it represents the time just
before breaking occurs in the experimental simulations (see
Fig. \ref{fig:full_experiment}).  As in the previous simulations, we
neglect the expansion potential ($\alpha_r = 0$) in $V_{ot}$
\eqref{eq:66}.

\begin{figure}
  \centering
  \begin{tabular}{cc}
    $\quad|\Psi|^2\qquad\qquad$ & $\qquad\qquad\qquad\quad(\arg \Psi)_r$
  \end{tabular}
  \includegraphics[scale=.45]{
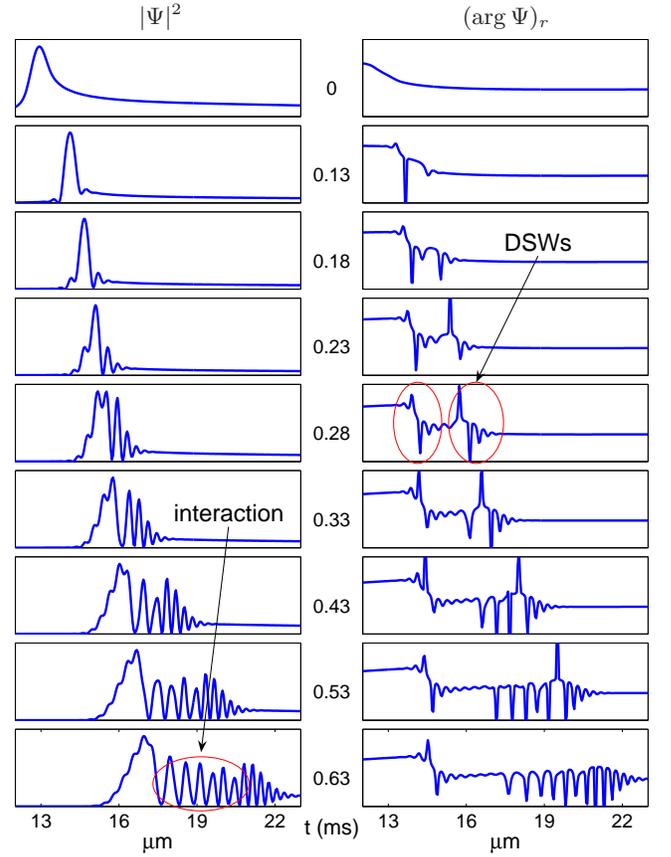}
  \caption{Three dimensional shock development and interaction for the
    modified out of trap experiment.  The left plot shows the
    evolution of the density $\abs{\Psi(r,0,t)}^2$ and the right plot
    shows the evolution of the radial velocity $[\arg \Psi(r,0,t)]_r$,
    both in the $z = 0$ plane.  As with the in trap experiment, the
    maximum absolute difference in density from the 2D evolution has a
    relative error of less than 1\%.}
  \label{fig:2d_3d_shock}
\end{figure}

In Fig. \ref{fig:2d_3d_shock}, the development and interaction of two
DSWs is shown from the evolution of the density
$|\Psi_{\text{3D}}(r,0,t)|^2$ and the radial velocity $[\arg
\Psi_{\text{3D}}(r,0,t)]_r$.  The signatures of two DSWs
(Fig. \ref{fig:gurevich}) are easier to see in the velocity plots on
the right of Fig. \ref{fig:2d_3d_shock} as the density modulations are
large.  At $t = 0.13$ ms, two DSWs begin to develop on the inner and
outer sides of the high density ring.  At $t = 0.33$ ms, these two
DSWs begin to interact giving rise to doubly periodic or multi-phase
behavior.  Following this, a DSW front propagates ahead of the
interaction region.  We will examine this interaction behavior in more
detail in the future.

In Fig. \ref{fig:1d_2d_shock}, the 3D and 1D simulations are compared
for the modified out of trap experiment.  The agreement is quite good
over the short time scale considered (1 ms).  This shows that the
\emph{initial} generation and interaction of DSWs in 3D BECs is well
explained by the 1D approximation.  At later times it is found that
the amplitudes and speeds diverge roughly similar to what we saw in
Fig. \ref{fig:1d_3d_shock_intrap}

\begin{figure}
  \centering
  \begin{tabular}{cc}
    $\qquad|\Psi|^2~\text{or}~\rho\quad$ &
    $\qquad\qquad\qquad\quad\eps(\arg 
    \Psi)_r~\text{or}~ u$
  \end{tabular}
  \includegraphics[scale=.45]{
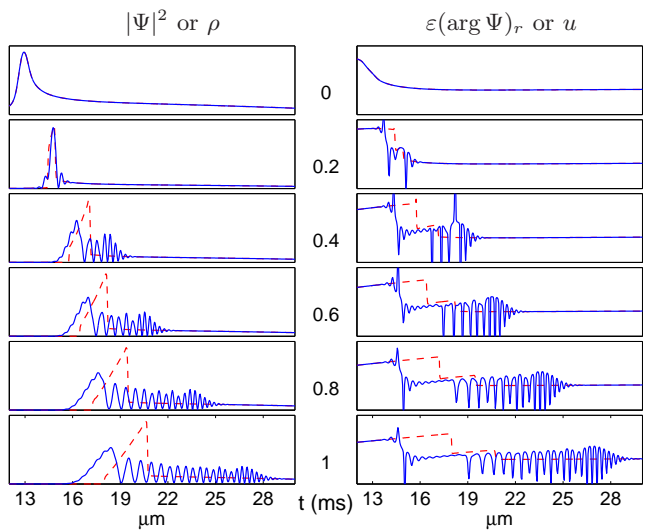}
  \caption{Shock evolution in 2D BEC (solid) and the dissipative
    regularization of the 2D Euler equations (dashed).  The left plot
    is density and the right plot is radial velocity.  The dissipative
    regularization was calculated using Clawpack \cite{Le02}.}
  \label{fig:2d_hydro_shock}
\end{figure}

Figure \ref{fig:2d_hydro_shock} shows a comparison of the dissipative
regularization of the 2D Euler equations \eqref{eq:118} and the
dispersive regularization given by the small dispersion limit of the
Gross-Pitaevskii equation \eqref{eq:116}.  The dissipative
regularization was calculated using Clawpack \cite{Le02}.  As in the
dispersive case, the gas dynamics equations develop two shock waves on
the inner and outer sides of the ring.  The inner shock wave
propagates as long as there is non-zero density on its inner edge.
This behavior is predicted by the 1D analysis where, with zero
background density, the solution is always a rarefaction (expansion)
wave.  The structure and speeds of the two types of shocks are quite
different as expected from the analysis given in this work.

\section{Conclusion}
\label{sec:conclusion}
We have presented new experimental evidence for dispersive shock waves
(DSWs) in a Bose-Einstein condensate.  Numerical simulations of the
experiments and comparisons with lower dimensional approximations show
that the experimentally observed ripples correspond to DSWs.  A DSW
has two associated speeds (front and rear of the oscillatory region)
and large amplitude oscillatory structure.

A detailed comparison between classical, dissipative shock waves and
dispersive shock waves has been undertaken in this work.  On a large
scale, a DSW has some similarities to a dissipative shock, but there
are fundamental and critical differences.  Using the notions of
averaging and weak limits, we have shown that a DSW has one behavior
similar to that of a classical shock--i.e. it has a steep front.
However, the key difference between a DSW and a dissipative shock is
in the shock speed and its structure. These properties are compared
using dissipative and dispersive regularizations of conservation laws.
A dispersive regularization for conservation laws gives rise to a
\emph{weak} limit in the sense that one must appropriately average
over the high frequency oscillations across a DSW.  On the other hand,
a dissipative regularization for conservation laws is a strong limit
which, in the case of a classical shock wave, converges to a
discontinuity propagating with a speed that satisfies the well known
jump conditions.

\begin{acknowledgments}
  This work was supported by NSF grants DMS-0303756 and VIGRE
  DMS-9810751.  The experimental part of this work was funded by NSF
  and NIST.
\end{acknowledgments}


\bibliographystyle{apsrev}

\end{document}